# Condensation in Dust-Enriched Systems

Denton S. Ebel[1]* and Lawrence Grossman[1,2]

[1]Department of the Geophysical Sciences, The University of Chicago, 5734 South Ellis Ave., Chicago, IL 60637

[2]Enrico Fermi Institute, The University of Chicago, 5640 South Ellis Ave., Chicago, IL 60637

* Corresponding author address:
Department of Earth and Planetary Sciences, American Museum of Natural History, New York, NY, 10024-5192
**E-mail:** debel@amnh.org

**Abstract** - Full equilibrium calculations of the sequence of condensation of the elements from cosmic gases made by total vaporization of dust-enriched systems were performed in order to investigate the oxidation state of the resulting condensates. The computations included 23 elements and 374 gas species and were done over a range of $P^{tot}$ from $10^{-3}$ to $10^{-6}$ bar and for enrichments up to 1000x in dust of C1 composition relative to a system of solar composition. Because liquids are stable condensates in dust-enriched systems, the MELTS non-ideal solution model for silicate liquids (Ghiorso and Sack, 1995) was incorporated into the computer code. Condensation at $10^{-3}$ bar and dust enrichments of 100x, 500x and 1000x occurs at oxygen fugacities of IW-3.1, IW-1.7 and IW-1.2, respectively, and, at the temperature of cessation of direct condensation of olivine from the vapor, yields $X_{Fa}$ of 0.019, 0.088 and 0.164, respectively. Silicate liquid is a stable condensate at dust enrichments >~12.5x at $10^{-3}$ bar and >~425x at $10^{-6}$ bar. At 500x, the liquid field is >1000K wide and accounts for a maximum of 48% of the silicon at $10^{-3}$ bar and is 240K wide and accounts for 25% of the silicon at $10^{-6}$ bar. At the temperature of disappearance of liquid, $X_{Fa}$ of coexisting olivine is 0.025, 0.14 and 0.31 at 100x, 500x and 1000x, respectively, almost independent of $P^{tot}$. At 1000x, the $Na_2O$ and $K_2O$ contents of the last liquid reach 10.1 and 1.3 wt%, respectively, at $10^{-3}$ bar but are both negligible at $10^{-6}$ bar. At $10^{-3}$ bar, iron sulfide liquids are stable condensates at dust enrichments at least as low as 500x and coexist with silicate liquid at 1000x. No sulfide liquid is found at $10^{-6}$ bar. At $10^{-3}$ bar, the predicted distribution of Fe between metal, silicate and sulfide at 1310K and a dust enrichment of 560x matches that found in H-Group chondrites, and at 1330K and 675x matches that of L-Group chondrites prior to metal loss.

Only at combinations of high $P^{tot}$ and high dust enrichment do the bulk chemical composition trends of condensates reach the FeO contents typical of Type IIA chondrules at temperatures where dust and gas could be expected to equilibrate, ≥ 1200K. Even under these conditions, however, the composition trajectories of predicted condensates pass through compositions with much more CaO + $Al_2O_3$ relative to MgO + $SiO_2$ than those of most Type IA chondrules. Furthermore, on a plot of wt% $Na_2O$ *vs*. wt% FeO, most chondrule compositions are too $Na_2O$-rich to lie along trends predicted for the bulk chemical compositions of the condensates at $P^{tot} \leq 10^{-3}$ bar and dust enrichments ≤ 1000x. Together, these chemical differences indicate that individual chondrules formed neither by quenching samples of the liquid + solid condensates that existed at various temperatures nor by quenching secondary liquids that formed from such samples. With the exception of very FeO-poor, $Na_2O$-rich glasses in Type I chondrules and glasses with very high FeO and $Na_2O$ in Type II chondrules, however, many chondrule glass compositions fall along bulk composition trajectories for liquids in equilibrium with cosmic gases at $10^{-3}$ bar and dust enrichments between 600x and 1000x. If these chondrules formed by secondary melting of mixtures of condensates that formed at different temperatures, nebular regions with characteristics such as these would have been necessary to prevent loss of $Na_2O$ by evaporation and FeO by reduction from the liquid precursors of their glasses, assuming that the liquids were hot for a long enough time to have equilibrated with the gas.

## 1. INTRODUCTION

Several lines of evidence suggest that most chondrites formed at oxygen fugacities significantly higher than those of a solar gas (*e.g.*, Fegley and Palme, 1985; Rubin *et al.*, 1988; Palme and Fegley, 1990; Weinbruch *et al.*, 1990). The most compelling evidence is the high FeO content of chondritic olivine and pyroxene grains, many of which have molar FeO/(FeO+MgO) ratios greater than 0.15 (Wood, 1967; Van Schmus, 1969). Grossman (1972) showed that the first olivine and pyroxene to condense from a cooling solar gas contain only trace amounts of FeO, because iron is more stable as co-condensing metallic nickel-iron. At equilibrium, olivine will not incorporate significant FeO until below 550 K, when iron metal reacts with gaseous $H_2O$ to form FeO (Grossman, 1972) which must then diffuse into the crystal structure of previously condensed forsterite, replacing MgO. At these low temperatures, however, this mechanism for producing the observed FeO content of olivine in chondrites encounters two fundamental problems: solid-gas equilibrium is unlikely, and diffusion in olivine is very slow. Enhancing the oxygen fugacity of the system in which chondritic matter formed is one way FeO could have been stabilized at temperatures high enough that it was incorporated into ferromagnesian silicates when, or soon after, they first condensed.

The most reasonable mechanism proposed for producing the oxygen fugacity required to form fayalitic olivine at higher temperatures is enhancement of the dust/gas ratio (Wood, 1967; Rubin *et al.*, 1988). In such a model, the initial nebula is a cold cloud of interstellar gas and dust, whose overall composition is solar and in which ~30% of the oxygen is in the dust, and virtually all of the H and C are in the gas. If, before nebular temperatures reach their maximum, dust concentrates in certain regions relative to the gas compared to solar composition, then total vaporization of such regions will produce a gas enriched in oxygen relative to hydrogen and carbon compared to solar composition. Subsequent condensation in such a region occurs in a gas with a significantly higher oxygen fugacity than one of solar composition. Furthermore, the abundance ratios of condensable elements such as Mg and Si to H are increased much more than the O/H ratio, because the dust contains nearly 100% of each of the condensable elements, compared to only 30% of the oxygen. The condensation temperature of any phase increases with increasing partial pressures of its gaseous constituents, which in turn increase with their abundances relative to hydrogen. Dust enrichment therefore not only increases oxygen fugacity, but also increases condensation temperatures, possibly to temperatures at which partial melts are stable.

Wood and Hashimoto (1993) and Yoneda and Grossman (1995) performed full equilibrium calculations of condensation in dust-enriched systems, and both studies found stability fields of silicate liquids at relatively low total pressure ($P^{tot}$). Therefore, an accurate thermodynamic description of silicate liquids is a prerequisite for an accurate description of condensation in dust-enriched systems. Yoneda and Grossman (1995) were the first to assess the stability of non-ideal $CaO$-$MgO$-$Al_2O_3$-$SiO_2$ (CMAS) silicate liquid (Berman, 1983), but were unable to address the stability of ferromagnesian liquids due to lack of an accurate thermodynamic model for silicate liquids containing Fe, Ti, Na and K.

The present work is the first to explore condensation in either solar composition or dust-enriched systems using a thermodynamic model for ferromagnesian liquids which has been tested against experimental data and natural assemblages. An 11-component subset of the 15-component "MELTS" silicate liquid model, developed by Ghiorso and Sack (1995) to model crystallization of natural silicate liquids of peridotite to intermediate compositions, has been incorporated into condensation calculations. In addition, this liquid model is shown here to describe accurately the crystallization of liquids in the FeO-CMAS system, similar to many of the liquids predicted in this work. Condensation sequences are computed at dust enrichments of up to 1000x, and at $P^{tot}$ of $10^{-3}$ and $10^{-6}$ bar, at temperatures from 1100 to 2400 K. Results indicate the composition changes in solid, liquid and gas phases likely to occur during direct condensation, partial evaporation, or pre-accretion metasomatism of matter in dust-enriched systems at these temperatures and pressures. The idea that ferromagnesian chondrules formed by direct condensation in the solar nebula has persisted since Sorby (1877) likened chondrules to solidified “drops of fiery rain”, and Wood (1967) revived it by suggesting that liquids of forsterite composition might be stable at low total pressures in gases enriched (by >5000x) in precondensed dust. Therefore, in this work, specific equilibrium assemblages are compared with specific chondrules, and the implications of dust enrichment for chondrule stability in the protoplanetary nebula are explored. Preliminary versions of this work were presented by Ebel and Grossman (1996, 1997a,b, 1998).

## 2. TECHNIQUE

### 2.1. Bulk Composition

The nature of the condensates from dust-enriched bulk compositions is strongly influenced by the composition assumed for the dust, and an infinite variety of fractionated dust compositions can be imagined. One constraint on dust composition, however, is that it lead to condensate assemblages containing chondritic proportions of condensable elements. C1 chondrites are representative of the bulk composition of the condensable fraction of solar system matter. If the bulk of the condensable elements was originally brought to the solar nebula in the form of interstellar dust, then it is reasonable to assume that the aggregate composition of that dust had a bulk chemical composition similar to that of C1 chondrites. Table 1 shows the relative atomic abundances of the 23 elements considered in this work in solar gas (Anders and Grevesse, 1989), the C1 chondrite dust component of solar gas, and several dust-enriched systems. For a dust enrichment of $n$, one way to calculate the bulk composition is by adding ($n$-1) units of the C1 dust to solar composition. Enrichment factors of up to 1000 were investigated. Although there are as yet no astronomical observations that confirm, or astrophysical models that produce, such enrichments, there exists no evidence to *rule out* such enrichments in protoplanetary environments.

**Table 1.** Relative atomic abundances in solar composition and the C1 component of solar composition, both normalized to $10^6$ atoms Si, and compositions of systems enriched in dust of C1 composition relative to solar.

| | Solar | Cl Dust | 100 x Cl | 500 x Cl | 1000 x Cl |
|---|---|---|---|---|---|
| H | $2.79 \times 10^{10}$ | $5.28 \times 10^{6}$ | $2.84 \times 10^{10}$ | $3.05 \times 10^{10}$ | $3.32 \times 10^{10}$ |
| He | $2.72 \times 10^{9}$ | | $2.72 \times 10^{9}$ | $2.72 \times 10^{9}$ | $2.72 \times 10^{9}$ |
| C | $1.01 \times 10^{7}$ | $7.56 \times 10^{5}$ | $8.50 \times 10^{7}$ | $3.87 \times 10^{8}$ | $7.65 \times 10^{8}$ |
| N | $3.13 \times 10^{6}$ | $5.98 \times 10^{4}$ | $9.05 \times 10^{6}$ | $3.30 \times 10^{7}$ | $6.28 \times 10^{7}$ |
| O | $2.38 \times 10^{7}$ | $7.63 \times 10^{6}$ | $7.80 \times 10^{8}$ | $3.83 \times 10^{9}$ | $7.65 \times 10^{9}$ |
| F | $8.43 \times 10^{2}$ | $8.43 \times 10^{2}$ | $8.43 \times 10^{4}$ | $4.22 \times 10^{5}$ | $8.43 \times 10^{5}$ |
| Ne | $3.44 \times 10^{6}$ | | $3.44 \times 10^{6}$ | $3.44 \times 10^{6}$ | $3.44 \times 10^{6}$ |
| Na | $5.74 \times 10^{4}$ | $5.74 \times 10^{4}$ | $5.74 \times 10^{6}$ | $2.87 \times 10^{7}$ | $5.74 \times 10^{7}$ |
| Mg | $1.07 \times 10^{6}$ | $1.07 \times 10^{6}$ | $1.07 \times 10^{8}$ | $5.37 \times 10^{8}$ | $1.07 \times 10^{9}$ |
| Al | $8.49 \times 10^{4}$ | $8.49 \times 10^{4}$ | $8.49 \times 10^{6}$ | $4.25 \times 10^{7}$ | $8.49 \times 10^{7}$ |
| Si | $1.00 \times 10^{6}$ | $1.00 \times 10^{6}$ | $1.00 \times 10^{8}$ | $5.00 \times 10^{8}$ | $1.00 \times 10^{9}$ |
| P | $1.04 \times 10^{4}$ | $1.04 \times 10^{4}$ | $1.04 \times 10^{6}$ | $5.20 \times 10^{6}$ | $1.04 \times 10^{7}$ |
| S | $5.15 \times 10^{5}$ | $5.15 \times 10^{5}$ | $5.15 \times 10^{7}$ | $2.58 \times 10^{8}$ | $5.15 \times 10^{8}$ |
| Cl | $5.24 \times 10^{3}$ | $5.24 \times 10^{3}$ | $5.24 \times 10^{5}$ | $2.62 \times 10^{6}$ | $5.24 \times 10^{6}$ |
| Ar | $1.01 \times 10^{5}$ | | $1.01 \times 10^{5}$ | $1.01 \times 10^{5}$ | $1.01 \times 10^{5}$ |
| K | $3.77 \times 10^{3}$ | $3.77 \times 10^{3}$ | $3.77 \times 10^{5}$ | $1.89 \times 10^{6}$ | $3.77 \times 10^{6}$ |
| Ca | $6.11 \times 10^{4}$ | $6.11 \times 10^{4}$ | $6.11 \times 10^{6}$ | $3.06 \times 10^{7}$ | $6.11 \times 10^{7}$ |
| Ti | $2.40 \times 10^{3}$ | $2.40 \times 10^{3}$ | $2.40 \times 10^{5}$ | $1.20 \times 10^{6}$ | $2.40 \times 10^{6}$ |
| Cr | $1.35 \times 10^{4}$ | $1.35 \times 10^{4}$ | $1.35 \times 10^{6}$ | $6.75 \times 10^{6}$ | $1.35 \times 10^{7}$ |
| Mn | $9.55 \times 10^{3}$ | $9.55 \times 10^{3}$ | $9.55 \times 10^{5}$ | $4.78 \times 10^{6}$ | $9.55 \times 10^{6}$ |
| Fe | $9.00 \times 10^{5}$ | $9.00 \times 10^{5}$ | $9.00 \times 10^{7}$ | $4.50 \times 10^{8}$ | $9.00 \times 10^{8}$ |
| Co | $2.25 \times 10^{3}$ | $2.25 \times 10^{3}$ | $2.25 \times 10^{5}$ | $1.13 \times 10^{6}$ | $2.25 \times 10^{6}$ |
| Ni | $4.93 \times 10^{4}$ | $4.93 \times 10^{4}$ | $4.93 \times 10^{6}$ | $2.47 \times 10^{7}$ | $4.93 \times 10^{7}$ |

## 2.2. Method of Calculation

The condensation code described here, "VAPORS", is described more completely by Ebel *et al.* (1999). All calculations are normalized to a total of one mole of atoms in the system. A typical condensation run at fixed $P^{tot}$ and bulk composition is begun with only the vapor phase present at 2400K. Most solutions are obtained at 10K intervals, using the result at the previous temperature as a first approximation. At each fixed pressure, temperature and bulk composition of the system, the partial pressures of the pure monatomic gaseous elements (the basis components of the gas phase) are obtained by calculating the distribution of the elements among 374 species in the gas phase, using standard techniques (Lattimer *et al*., 1978; Smith and Missen, 1982). The stability of each potential, stoichiometrically pure, single component condensate phase is then evaluated from the partial pressures of the elements and the Gibbs energy for that phase, by considering the energy balance of the formation reaction of the condensate from the monatomic gaseous elements. In the case of a liquid or solid solution phase, the "best" composition is determined by finding that composition at which the activities of the components describing the solution phase most closely match equivalent activities in the gas, using the algorithms of Ghiorso (1994). This composition is then tested for stability in much the same way as a stoichiometrically pure condensate, but also accounting for the thermodynamic mixing properties of the components in the solution phase. In some cases where silicate liquid is present, this algorithm failed to find the “best” pyroxene solid solution composition, and the program proceeded with a pure diopside end-member composition instead. In such cases, the program was restarted with a “seed”, Ti, Al-bearing diopsidic pyroxene substituted for the pure diopside at and above the temperature step at which pure diopside had been found to be stable. In all such cases, a complex, Ti-, Al-bearing diopsidic pyroxene was found to be stable at the temperature where pure diopside had been found, or, at most, 20K higher. The Gibbs energy of the system was always on the order of 0.5 J lower per mole of elements in the system for the assemblage with the pyroxene solid solution than for the one with pure diopside. This problem occurs only with the pyroxene solid solution model, probably because of the difficulty in determining both the composition and ordering state of the near end-member pyroxene in equilibrium with a gas phase highly depleted in some of the pyroxene-forming elements. Once a phase is determined to be stable, it is added to the stable assemblage in a seed amount ($10^{-7}$ moles), which is subtracted from the gas. The next step is to distribute mass between the phases to minimize the total free energy of this new system.

In this work, the second order technique of Ghiorso (1985), following Betts (1980), was adapted to the problem of distributing mass between phases to minimize directly the total Gibbs free energy of a system consisting of gas and multiple pure and solution phases, both solid and liquid. The Gibbs energy of the entire system can be imagined as a surface in $m$ dimensions, where $m$ is the total number of components independently variable in each of the phases present. The components of the gas are the monatomic elements, while those of solution phases are the endmembers of these phases. Each distribution of elements between these components at fixed temperature and pressure defines a thermodynamic state of the system and corresponds to a point on the Gibbs surface. In successive iterations, information about the local slope and curvature of the Gibbs surface at the current state of the system is used to determine the direction toward a minimum on this surface, along which the next iterative solution must lie. Then atoms are redistributed accordingly among the gas and condensates, that is among the *m* components, so that this minimum is approached as closely as possible. From the perspective of this new state of the system, the Gibbs surface "looks" different, so a new minimum must be sought in a further iteration. Convergence is declared when the vector norm of all the changes in composition in the *m* directions does not change by > $10^{-12}$ between iterations. The VAPORS program usually converges in less than ten iterations in this part of the algorithm.

Upon convergence to a free energy minimum, the stabilities of non-condensed phases are assessed as described above, and if additional phases are found to be stable relative to the gas, they are added as described above and the minimization algorithm is repeated. Even trace phases such as perovskite are typically present at levels >$10^{-6}$ moles per mole of elements in the complete system. If the amount of a phase drops below a minimum value, set at $10^{-10}$ moles, that phase is removed from the condensate assemblage, and the minimization algorithm is repeated. If no phase must be added or removed after the minimization, the system is considered solved for that temperature, pressure and bulk composition, and a new temperature step is initiated.

Convergence of each solution is assessed independently by calculation of the difference in the chemical potential of each condensed component between the gas and condensates. For temperatures >1400K, these differences for each component are always < $10^{-7}$ of the chemical potential in the gas, and usually very much better, e.g., ~$10^{-12}$. At lower temperatures, particularly in dust-enriched systems, these differences in some cases increase for components containing the elements Ca, Al, and Ti, and no results are reported here for any temperature step in which the difference exceeds $10^{-4}$ for any condensate component. Even in an example where these differences are ~$3x10^{-4}$, they would record uncertainties corresponding to a shift of only ~$10^{-10}$ of the total Ca in the system between the gas and the condensate assemblage. These reaction imbalances occur because the algorithms call for numerical approximation of the first and second derivatives of the Gibbs energy of the gas with respect to the concentration of each of the condensing elements in it, and this approximation becomes increasingly sensitive to machine numerical precision at very low concentrations of elements in the gas (*e.g.*, $10^{-20}$ moles per mole of elements in the system). Mass balance is preserved to within <$10^{-27}$ of the moles of atoms present throughout all calculations.

**Table 2.** Gas species and thermodynamic data sources included in the calculation, in addition to species used by Grossman (1972) and/or Yoneda and Grossman (1995).

| Chase *et al.* (1985) | | | | | Other Sources |
|---|---|---|---|---|---|
| Ne | $CHClF_2$ | NiCl | $CoF_2$ | $F_3NO$ | K-1991[b] |
| Ar | $CHCl_2F$ | ClO | CrN | $PF_3O$ | $CrCl_2O_2$ |
| AlClF | $CHCl_3$ | TiOCl | CrO | $PF_3$ | $MnCl_2$ |
| $AlClF_2$ | CHF | $ClO_2$ | $CrO_2$ | $F_3PS$ | CrS |
| $AlCl_2$ | CHFO | PCl | $CrO_3$ | $SF_3$ | $MnF_2$ |
| $AlCl_2F$ | $CHF_3$ | ClS | $F_{10}S_2$ | $SiF_3$ | $NiF_2$ |
| $AlCl_3$ | $CH_2ClF$ | $ClS_2$ | FeF | $TiF_3$ | $Ni(OH)_2$ |
| $AlF_2$ | $CH_2Cl_2$ | SiCl | FHO | $H_4F_4$ | NiO |
| $AlF_2O$ | $CH_2F_2$ | TiCl | $FHO_3S$ | $Mg_2F_4$ | $NiAl_2Cl_8$ |
| $AlF_3$ | $CH_3Cl$ | $Cl_2$ | $SiFH_3$ | $N_2F_4$ | NiF |
| $NaAlF_4$ | $CH_3F$ | Co $Cl_2$ | $SiF_3H$ | $SF_4$ | NiH |
| AlHO | KCN | $Cl_2FOP$ | FNO | $SiF_4$ | TiS |
| $Al_2$ | $CN_2$[a] | $FeCl_2$ | $FNO_2$ | $TiF_4$ | $CrCl_2$ |
| $Al_2Cl_6$ | CP | $SiCl_2H_2$ | $FNO_3$ | TiFO | |
| $Al_2F_6$ | $C_2Cl_2$ | $K_2Cl_2$ | FO | $PF_5$ | |
| CCl | $C_2Cl_4$ | $C_2Cl_4$ | $Na_2Cl_2$ | $H_5F_5$ | PM-1983[b] |
| CClFO | $C_2Cl_6$ | $NiCl_2$ | $FO_2$ | $H_6F_6$ | |
| $CClF_3$ | $C_2F_2$ | $Cl_2O$ | FPS | $SF_6$ | CoO |
| CClN | $C_2F_4$ | $TiCl_2O$ | SF | $H_7F_7$ | MnO |
| CClO | $C_2F_6$ | $Cl_2O_2S$ | SiF | FeS | |
| $CCl_2$ | $C_2HCl$ | $Cl_2S$ | $F_2$ | $HNO_2$[a] | |
| $CCl_2F_2$ | $C_2HF$ | $Cl_2S_2$ | $FeF_2$ | $K_2O_2H_2$ | |
| $CCl_2O$ | $C_2H_4$ | $SiCl_2$ | $H_2F_2$ | KO | |
| $CCl_3$ | $(KCN)_2$ | $TiCl_2$ | $K_2F_2$ | $K_2$ | |
| $CCl_3F$ | $Ni(CO)_4$ | $CoCl_3$ | $F_2N$ | $K_2SO_4$ | |
| $CCl_4$ | $Fe(CO)_5$ | $SiCl_3F$ | $F_2N_2$[a] | $Na_2SO_4$ | |
| CF | CaCl | $FeCl_3$ | $Na_2F_2$ | NiS | |
| CFN | CaS | $SiCl_3H$ | $F_2O$ | $P_4O_{10}$ | |
| CFO | CoCl | $POCl_3$ | $SiF_2O$ | $P_4O_6$ | |
| $CF_2$ | ClF | $PCl_3$ | $TiF_2O$ | $P_4$ | |
| $CF_2O$ | $ClFO_2S$ | $PCl_3S$ | $SF_2O_2$ | $P_4S_3$ | |
| $CF_3$ | $ClFO_3$ | $TiCl_3$ | $PF_2$ | $S_3$ | |
| $CF_4$ | $ClF_2OP$ | $Co_2Cl_4$ | $SF_2$ | $S_4$ | |
| $CF_4O$ | $ClF_3$ | $Fe_2Cl_4$ | $S_2F_2$[a] | $S_5$ | |
| $CF_8S$ | $ClF_5$ | $Mg_2Cl_4$ | $SiF_2$ | $S_6$ | |
| CHCl | $ClF_5S$ | $SiCl_4$ | $FeF_3$ | $S_7$ | |
| $ClNO_2$ | FeCl | $TiCl_4$ | $SiF_2H_2$ | | |
| ClHO | $PCl_5$ | $SiCl_3$ | $H_3F_3$ | | |
| ClNO | $Fe_2Cl_6$ | $SiCl_3F$ | $NF_3$ | | |

[a] Both *cis* and *trans* forms are included.

[b] K-1991 = Knacke et al. (1991); PM-1983 = Pedley and Marshall (1983).

## 2.3. Thermodynamic Data for Elements and Gas Species

In each calculation, 23 elements were included: H, He, C, N, O, F, Ne, Na, Mg, Al, Si, P, S, Cl, Ar, K, Ca, Ti, Cr, Mn, Fe, Co and Ni. The gas species considered in every calculation include all species considered by Grossman (1972) and Yoneda and Grossman (1995), as well as those listed in Table 2. The $\Delta_f H^\circ$(298.15K), $S^\circ$(298.15K) and $C_p^\circ(T)$ data for gas species and elements in their standard states were taken wherever possible from the JANAF tables (Chase *et al*., 1985), obtained in machine readable form from the National Institute of Standards and Technology in 1995. For a few gas species not present in the JANAF database, data from Knacke *et al*. (1991) or Pedley and Marshall (1983) were used. During calculation, apparent Gibbs energies of formation (Anderson and Crerar, 1993), and hence the equilibrium constants of reactions, were calculated by integration of polynomial fits to tabulated $C_p^\circ(T)$ data. Errors were found in the JANAF tabulations (Chase *et al*., 1985) of the Gibbs energy of formation ($\Delta_f G^\circ$)and equilibrium constant ($\ln K_f$) for the gas species $C_2N_2$, $C_2H_2$, CN, and HS. The error in HS was also present in the electronic version, and had not been previously reported (M. Chase, pers. comm; 1996). Although it does not occur in the tabulations of Stull and Prophet (1970), Barin (1989), or Knacke *et al*. (1991), the $HS_{(g)}$ error has been propagated through the work of Yoneda and Grossman (1995), and probably also Sharp and Wasserburg (1995) and others. The effect of this error is to overestimate the stability of $HS_{(g)}$, and cause $SiS_{(g)}$ to sequester slightly less Si than it should.

## 2.4. Thermodynamic Data and Models for Solids

The internally consistent thermodynamic database of Berman (1988), or a combination of the internally consistent databases of Berman and Brown (1985) and Berman (1983) were used wherever possible for all potential condensates in Table 3 and for most end-member components of the solid solution series in Table 4, except for the metal alloy. This means that Berman (1988) was the source of end-member data for the melilite and feldspar solid solutions, not the references cited for the solution models for these phases. The JANAF data (Chase *et al*., 1985) for pyrrhotite, $Fe_{0.877}S$, are based on estimation of heat capacities from 600 to 1475K. Recent work below 1000K by Grønvold and Stølen (1992) indicates that these data cause over-stabilization of pyrrhotite by ~5 kJ at 1000K. Therefore, Gibbs energies of formation of pyrrhotite from the JANAF tables were revised upward by this amount in the calculation. This revision lowers the appearance temperature of pyrrhotite by ~50K, compared to the JANAF data.

**Table 3.** Pure solid phases considered in the calculation and sources of thermodynamic data.

Miscellaneous solid phases[a]

| | | |
|---|---|---|
| Aenigmatite | $Na_2Fe_5TiSi_6O_{20}$ | M |
| Andalusite | $Al_2SiO_5$ | B8 |
| Anhydrite | $CaSO_4$ | R |
| Anthophyllite | $Mg_7Si_8O_{22}(OH)_2$ | B8 |
| Apatite | $Ca_5(PO_4)_3OH$ | M |
| Brucite | $Mg(OH)_2$ | B8 |
| Ca-aluminate | $CaAl_2O_4$ | B5 |
| Calcite | $CaCO_3$ | B8 |
| Cohenite | $Fe_3C$ | R |
| Cordierite | $Mg_2Al_4Si_5O_{18}$ | B8 |
| Corundum | $Al_2O_3$ | B8 |
| Cristobalite | $SiO_2$ | M |
| Dolomite | $CaMg(CO_3)_2$ | B8 |
| Grossite | $CaAl_4O_7$ | B5 |
| Hatrurite | $Ca_3SiO_5$ | B5 |
| Hibonite | $CaAl_{12}O_{19}$ | B3 |
| Kalsilite | $KAlSiO_4$ | M |
| Leucite | $KAlSi_2O_6$ | M |
| Lime | $CaO$ | B8 |
| Magnesite | $MgCO_3$ | B8 |
| Manganosite | $MnO$ | R |
| Merwinite | $Ca_3MgSi_2O_8$ | B8 |
| Nepheline | $NaAlSiO_4$ | M |
| Periclase | $MgO$ | B8 |
| Perovskite | $CaTiO_3$ | R |
| Pyrrhotite | $Fe_{0.877}S$ | J |
| Quartz | $SiO_2$ | M |
| Rankinite | $Ca_3Si_2O_7$ | B5 |
| Rutile | $TiO_2$ | B8 |
| Sapphirine | $Mg_4Al_{10}Si_2O_{23}$ | B3 |
| Sillimanite | $Al_2SiO_5$ | B8 |
| Sinoite | $Si_2N_2O$ | F |
| Sphene | $CaTiSiO_5$ | B8 |
| Talc | $Mg_3Si_4O_{10}(OH)_2$ | B8 |
| Tialite | $Al_2TiO_5$ | R |
| Tri-Ca aluminate | $Ca_3Al_2O_6$ | B5 |
| Tridymite | $SiO_2$ | M |
| Troilite | $FeS$ | C |
| Whitlockite | $Ca_3(PO_4)_2$ | M |
| Wollastonite | $CaSiO_3$ | B8 |

Chase *et al.* (1985)

| | |
|---|---|
| Al | MgS |
| $Al_4C_3$ | $Mg_2Si$ |
| AlN | $MgSO_4$ |
| $Al_2S_3$ | $MgTi_2O_5$ |
| $Al_6Si_2O_{13}$ | Na |
| alpha Ca | alpha $Na_3AlF_6$ |
| beta Ca | beta $Na_3AlF_6$ |
| $CaCl_2$ | $NaAlO_2$ |
| $CaF_2$ | NaCl |
| $Ca(OH)_2$ | NaCN |
| CaS | $Na_2CO_3$ |
| CoO | NaF |
| $Cr_3C_2$ | NaH |
| CrN | $NaO_2$ |
| $Cr_2N$ | $Na_2O$ |
| $Cr_2O_3$ | $Na_2O_2$ |
| $FeCl_2$ | NaOH |
| $FeF_2$ | $Na_2S$ |
| $Fe_{0.947}O$ | $Na_2S_2$ |
| FeO | $Na_2SiO_3$ |
| $Fe(OH)_2$ | $Na_2Si_2O_5$ |
| $Fe(OH)_3$ | $Na_2SO_4$(I-V) |
| $FeS_2$ (Pyrite) | $NH_4Cl$ |
| $FeSO_4$ | P |
| $Fe_2(SO_4)_3$ | monocl S |
| Graphite | ortho S |
| K | alpha SiC |
| KCl | beta SiC |
| KF | $Si_3N_4$ |
| $KF_2H$ | $SiS_2$ |
| KH | alpha Ti |
| $K_2O$ | beta Ti |
| KOH | TiC |
| $K_2S$ | $TiH_2$ |
| $K_2SO_4$ | TiN |
| $K_2SiO_3$ | alpha TiO |
| Mg | beta TiO |
| $MgC_2$ | $Ti_2O_3$ |
| $Mg_2C_3$ | $Ti_4O_7$ |
| $MgCl_2$ | alpha $Ti_3O_5$ |
| $MgH_2$ | beta $Ti_3O_5$ |
| $MgF_2$ | |
| $Mg_3N_2$ | |

------------------------------------------------------------------------------------------------------------

**Table 3:** (notes)

[a] Symbols for data are: B5=Cp from Berman and Brown (1985), 298 K data from Berman (1983); B3=Berman (1983); B8=Berman (1988); C=Hsieh *et al.* (1987); R=Robie *et al.* (1978); F=Fegley (1981); M='MELTS' software database (Ghiorso & Sack, 1995); J= Chase *et al.* (1985) modified for consistency with Grønvold & Stølen (1992).

**Table 4.** Solid solutions considered in the calculation, and sources of solution models.

| | |
|---|---|
| Metal alloy (this work) | |
| Iron | Fe |
| Nickel | Ni |
| Silicon | Si |
| Chromium | Cr |
| Cobalt | Co |
| Olivine (Sack and Ghiorso, 1989, 1994b) | |
| Fayalite | $Fe_2SiO_4$ |
| Forsterite | $Mg_2SiO_4$ |
| Monticellite | $CaMgSiO_4$ |
| Melilite (Charlu *et al.*, 1981) | |
| Åkermanite | $Ca_2MgSi_2O_7$ |
| Gehlenite | $Ca_2Al_2SiO_7$ |
| Orthopyroxene (Sack & Ghiorso, 1989, 1994b) | |
| Enstatite | $Mg_2Si_2O_6$ |
| Ferrosilite | $Fe_2Si_2O_6$ |
| Ca-pyroxene (Sack & Ghiorso, 1994a,b,c) | |
| Diopside | $CaMgSi_2O_6$ |
| Hedenbergite | $CaFeSi_2O_6$ |
| Alumino-buffonite | $CaTi_{0.5}Mg_{0.5}AlSiO_6$ |
| Buffonite | $CaTi_{0.5}Mg_{0.5}FeSiO_6$ |
| Essenite | $CaFeAlSiO_6$ |
| Jadeite | $NaAlSi_2O_6$ |
| Feldspar (Elkins and Grove, 1990) | |
| Albite | $NaAlSi_3O_8$ |
| Anorthite | $CaAl_2Si_2O_8$ |
| Sanidine | $KAlSi_3O_8$ |
| Spinel (Sack and Ghiorso, 1991a,b) | |
| Chromite | $FeCr_2O_4$ |
| Hercynite | $FeAl_2O_4$ |
| Magnetite | $Fe_3O_4$ |
| Spinel | $MgAl_2O_4$ |
| Ulvospinel | $Fe_2TiO_4$ |
| Rhombohedral oxide (Ghiorso, 1990) | |
| Geikielite | $MgTiO_3$ |
| Hematite | $Fe_2O_3$ |
| Ilmenite | $FeTiO_3$ |
| Pyrophanite | $MnTiO_3$ |

The solid solution models implemented in the MELTS program (*circa* 1993; Table 4) were used in all calculations, except that Ca-pyroxenes (Sack and Ghiorso, 1994a,b,c) were constrained to have 1 total atom of Ca + Na per 6 oxygen atoms. These represent the most comprehensive treatments of the anhydrous igneous rock-forming minerals presently available and are the solid solution models against which the MELTS silicate liquid model is calibrated. In addition, solid Fe-Ni-Si-Cr-Co alloy was modeled using JANAF data (Chase *et al*., 1985) for pure metal endmembers, and an asymmetric binary solution model calibrated against activity data for the binary systems of Chuang *et al*. (1986b) for Fe-Ni, Sakao and Elliott (1975) for Fe-Si, and Normanton *et al*. (1976) for Fe-Cr, with Fe-Co treated as ideal. Such a calibration is justifiable for the dilute alloys found at high temperature in this work.

Some cations of great interest in condensation are not contained in some of the liquid or solid solution models used here. These are the first condensation calculations in which the $TiO_2$ content of spinel is modeled, and extraordinarily high $TiO_2$ contents are predicted at very high temperatures. In all such cases, however, spinel coexists with a CMAS liquid into which $TiO_2$ is artificially prevented from dissolving. Partitioning experiments (Connolly and Burnett, 1999) suggest that these high $TiO_2$ contents may be spurious. Insufficient experimental work exists to justify inclusion of $Ti^{3+}$ or $Cr^{3+}$ in the pyroxene model. No solution model is used for Mn, S, P or C in the metal alloy, and this could artificially enhance the stabilities of troilite, pyrrhotite and whitlockite. Similarly, our inability to account for Ni or Co in troilite or pyrrhotite, nor for Cr, Ti or Al in olivine, may artificially destabilize these phases slightly. Although Hirschmann (1991) has modeled Ni, Co and Mn in olivine, these elements are not addressed by the pyroxene model, nor are Ni and Co included in the spinel model employed here. Because inclusion of Ni, Co or Mn in only one of these phases would artificially stabilize that phase and cause it to contain excessive amounts of these cations, these cations were not included in the olivine model. This omission, however, artificially stabilizes $MnTiO_3$-rich rhombohedral oxide solid solutions and crystalline MnO.

### 2.5. Thermodynamic Data and Models for Silicate Liquids

A major innovation in the work presented here is the inclusion of the "MELTS" model for silicate liquids (Ghiorso, 1985; Ghiorso and Sack, 1995), which describes the thermodynamic properties of silicate liquids using a regular (symmetric) binary solution model in the components $SiO_2$-$TiO_2$-$Al_2O_3$-$Fe_2O_3$-$Fe_2SiO_4$-$Mg_2SiO_4$-$MgCr_2O_4$-$CaSiO_3$-$Na_2SiO_3$-$KAlSiO_4$-$H_2O$, in addition to $MnSi_{0.5}O_2$-$NiSi_{0.5}O_2$-$CoSi_{0.5}O_2$-$Ca_3(PO_4)_2$ which have been omitted in the present study. Crystallization calculations with MELTS have been found to yield remarkable agreement between calculated and observed amounts and compositions of phases in liquid-crystal equilibrium experiments at 1 bar (Ghiorso and Carmichael, 1985) and at 10 kbar (Baker *et al*., 1995; Hirschmann *et al.*, 1998). Ghiorso and Sack (1995) caution against using their model (a) in systems containing only a small subset (<7) of the components, or (b) far outside the temperature-pressure-composition range of its calibration. Both of these *caveats* are addressed below.

**Table 5.** Comparison of experimental phase appearance temperatures (Takahashi, 1986, run 4) with MELTS calculations for KLB-1, all at 1 bar.

| | **Temperature (K)** | |
|---|---|---|
| | **reported** | **predicted by MELTS** |
| Olivine in: | >1973 | 1993 |
| Spinel in: | 1723-1773 | 1838 |
| Orthopyroxene in: | 1573-1623 | 1498 |
| Ca- rich pyroxene in: | 1473-1523 | 1473 |
| Feldspar in: | 1423-1448 | 1463 |
| Liquid out: | 1373-1423 | 1368 |

*2.5.1. Test of* MELTS*: Peridotite* KLB-1

Anticipating that condensate liquids will be poor in non-CMAS components and will thus violate *caveat* (a), we tested MELTS calculations against quenched partial melting experiments of peridotite KLB-1, whose non-CMAS components consist of only 8.1 wt% FeO, and ≤ 0.3% of all other oxides. Takahashi (1986) and Takahashi *et al.* (1993) reported the temperature intervals between the observed absence and presence of phases, as well as phase compositions and melt fractions for KLB-1 at 1 bar at the Ni-NiO oxygen buffer. Note that only one of their seven data points used here is used in the MELTS calibration database. It can be seen in Fig. 1 that the MELTS model reproduces the observed volume fractions of liquids well, except at low melt fractions where there may be significant measurement error in the experiments. The solidus temperature and appearance temperatures of olivine, Ca-pyroxene, and feldspar agree nearly within experimental error, but the model underpredicts the crystallization temperature of orthopyroxene and overpredicts that of spinel (Table 5). Hirschmann *et al.* (1998) observed that the MELTS model *over*predicted the crystallization temperature of orthopyroxene at 10 kbar. These differences reflect compromises made by Sack and Ghiorso (1994c) to best satisfy both high- and low-pressure pyroxene-liquid phase relations. In Fig. 2, the 1 bar liquid compositions are compared with MELTS results, with all $Fe_2O_3$ recalculated to FeO. The good agreement of the results for melt fraction and composition suggests that the MELTS model will yield reasonably accurate results in the condensation calculation, particularly because olivine dominates the distribution of mass in condensation sequences. Because spinel is a minor phase, overstabilization of spinel will not have a significant effect on liquid stability. The understabilization of orthopyroxene, relative to liquid, suggests that liquid stability might be slightly overpredicted when orthopyroxene condenses with it, and that the temperature of appearance of the latter phase in the condensation calculation may be too low.

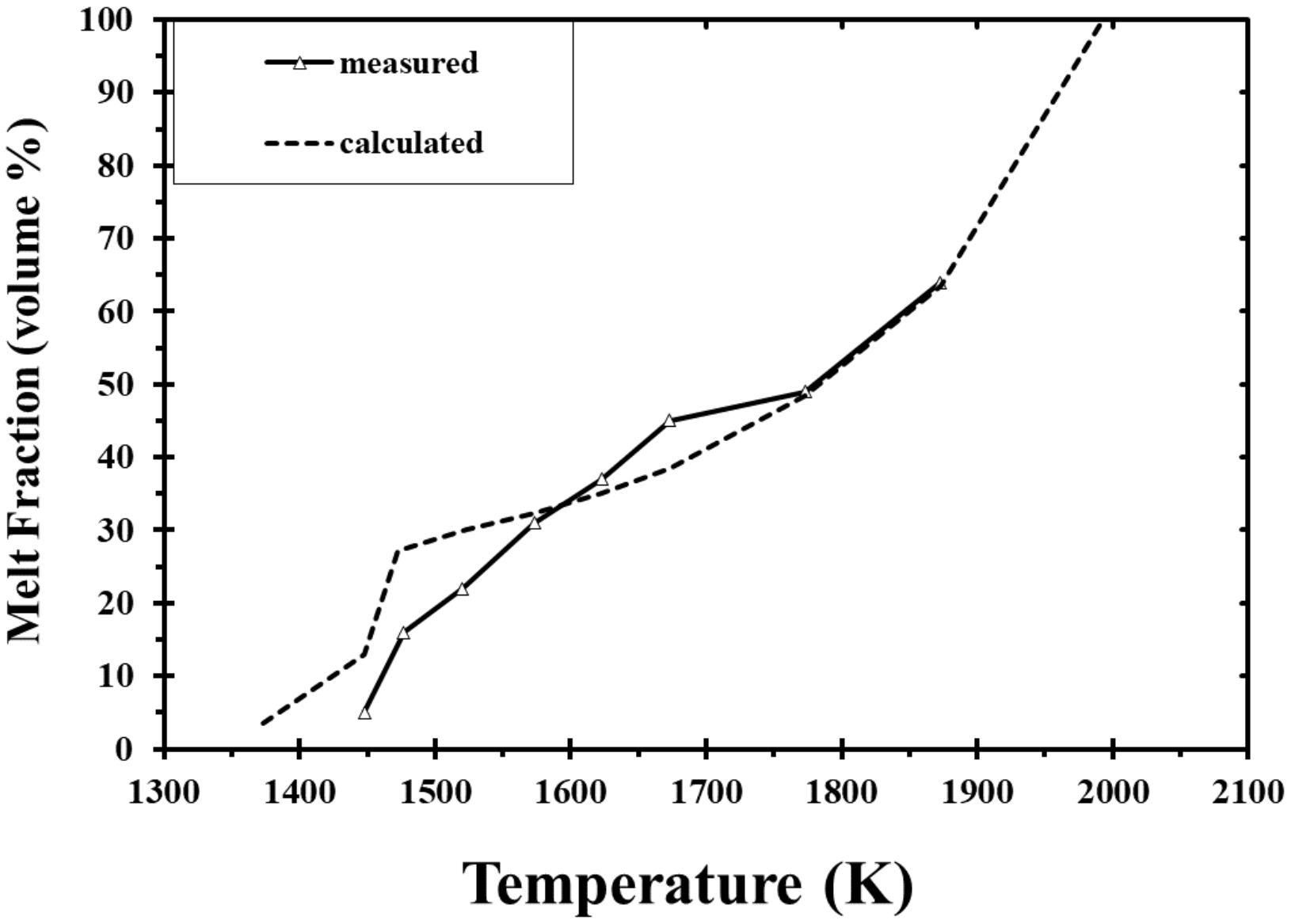

**Figure 1:** Comparison of melt fractions measured in peridotite melting experiments of Takahashi (1986) and Takahashi *et al*. (1993) with those calculated at 1 bar using MELTS.

In addition to MELTS, Berman's (1983) model for CaO-MgO-$Al_2O_3$-$SiO_2$ (CMAS) liquids is included in the present work. Yoneda and Grossman (1995) used this model, and explained in detail its advantages and drawbacks. The CMAS liquid model works well at high temperatures, where these four oxides are the only major ones condensed, but it is inadequate under conditions where FeO, $Na_2O$ and other non-CMAS components condense in appreciable quantities. Therefore, the MELTS model must be used at lower temperatures where non-CMAS oxides are important constituents of the liquid. The purely CMAS liquid region is very far outside the composition range over which MELTS is calibrated and contains too few components for reliable application of the MELTS liquid model. Furthermore, because the MELTS liquid uses mostly silicate components, not pure oxides, as endmembers, it cannot be applied to some especially Ca- and Al-rich regions of composition space that are treated adequately by the CMAS liquid model. For example, $CaSiO_3$ is the major Ca-containing component employed by MELTS; yet, early high temperature condensate liquids never contain as much $SiO_2$ as CaO. For these reasons, both models are required to completely describe condensation of silicate liquids over the temperature ranges where liquids may be stable.

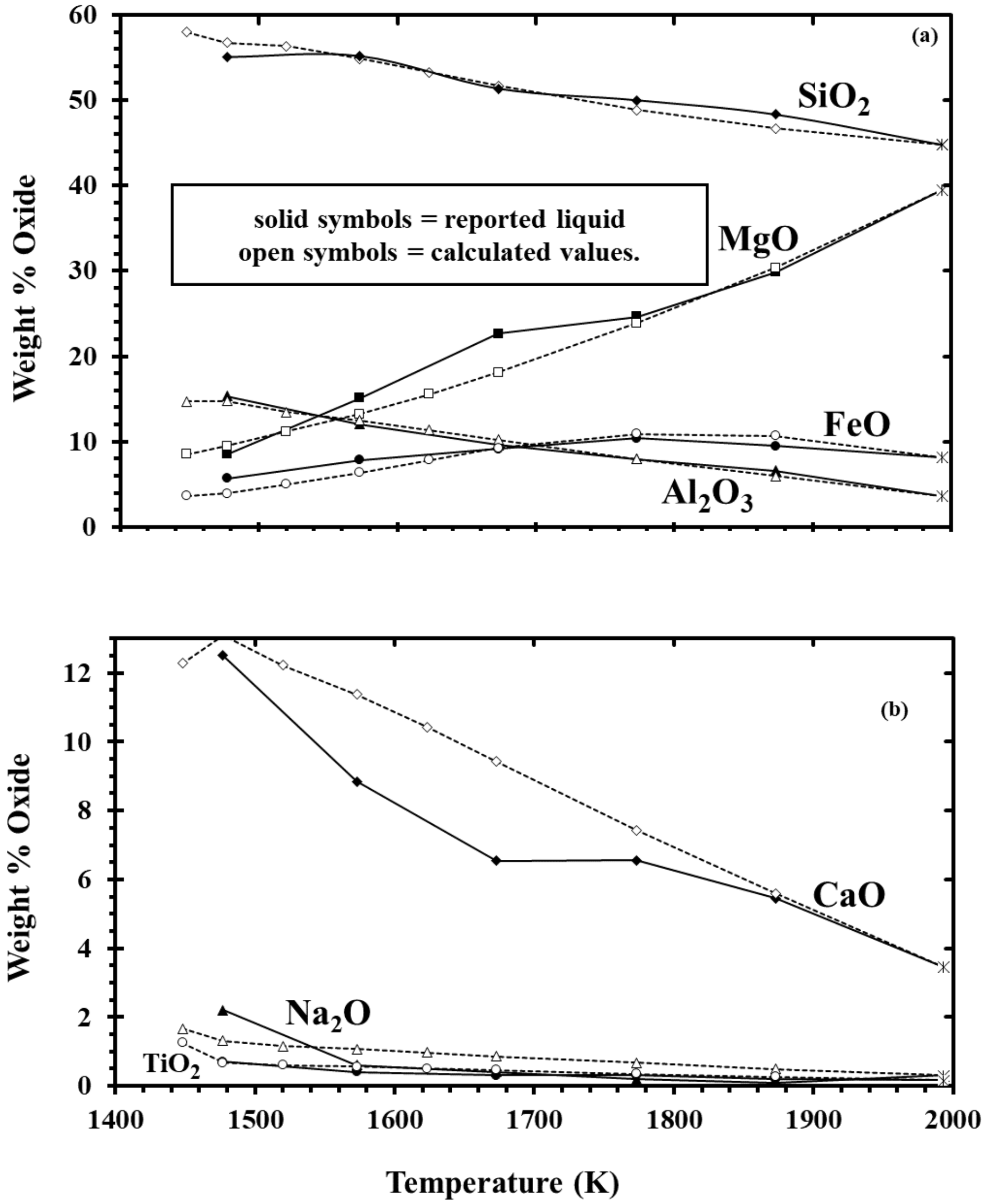

**Figure 2:** Comparison of measured compositions of KLB-1 liquids (Takahashi, 1986; Takahashi *et al.*, 1993) with those calculated from MELTS, all at one bar. Asterisks indicate starting compositions in the experiments.

*2.5.2. Transition Between Liquid Models*

A decision must be made as to when to switch from one model to the other. In order to model non-CMAS oxides in the liquid, it would be best to switch to the MELTS model at the highest feasible temperature. For the case of 100x dust enrichment at $P^{tot}=10^{-3}$ bar, the curves in Fig. 3 illustrate the major oxide compositions of the two liquids, calculated at 2K intervals, near the appearance temperature of olivine, indicated by the vertical line at 1782K. The CMAS liquid is CaO- and $Al_2O_3$-rich at high temperatures, but $SiO_2$ and MgO increase rapidly with decreasing temperature. By contrast, although a MELTS liquid becomes stable well above 1782K, it is CaO-deficient and $SiO_2$-enriched, relative to the CMAS liquid, because the only liquid the MELTS model can determine to be stable must have sufficient $SiO_2$ to supply the required $CaSiO_3$ component. That is, in the temperature range above at least 1790K, the most stable liquid possible in the MELTS composition range is not the liquid which *should* be stable. When the temperature of olivine appearance is reached, however, the MELTS liquid has gained sufficient $SiO_2$ and MgO to have a composition very similar to the CMAS liquid at the same temperature. Once sufficient $SiO_2$ has condensed, the MELTS model closely tracks the Berman (1983) model liquid, but also accounts for increasing FeO and $TiO_2$ contents.

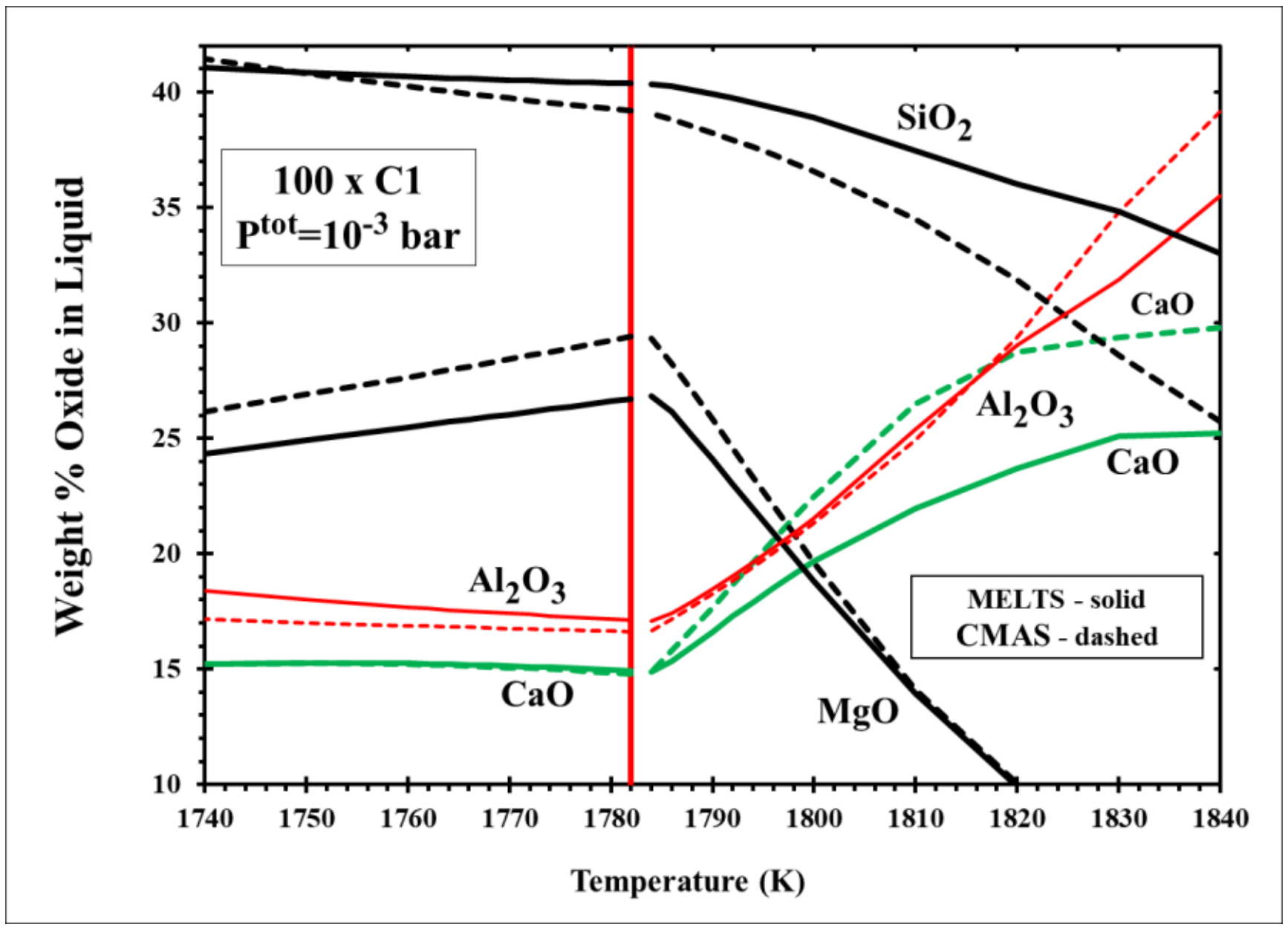

**Figure 3:** Compositions of CMAS and MELTS liquids near the olivine appearance temperature (vertical line at 1782 K) at the stated conditions. In this temperature range, the MELTS liquid also contains ~1 wt% of other oxides, which are not shown.

It was determined by performing condensation calculations with each liquid separately that the CMAS liquid and the MELTS liquid have nearly identical compositions at the temperature where olivine becomes stable with CMAS liquid for $10^{-6} \leq P^{tot} \leq 10^{-3}$, and 15x ≤ dust enrichment ≤ 1000x. The criterion of olivine stability is, therefore, used to trigger a switch from the CMAS liquid model to the MELTS silicate liquid model in the calculations. The similarity in oxide concentrations below the olivine stability temperature (Fig. 3) could be expected from the similarity of the Berman (1983) database, against which the CMAS model was calibrated, and the Berman (1988) database, upon which the MELTS model relies. These comparisons strongly suggest that the *caveats* cited by Ghiorso and Sack (1995) regarding use of the MELTS liquid model are not egregiously violated in its use below the condensation temperature of olivine.

## 3. RESULTS

### 3.1. Vapor of Solar Composition

Different thermodynamic data are employed for some crystalline phases, many more chemical species are included and a very different computational procedure was used in the present study than in our previous work on condensation (Yoneda and Grossman, 1995). It is therefore important to compare results from the two studies, and this is done for the case of a solar gas at $P^{tot} = 10^{-3}$ bar in Table 6. Appearance temperatures of phases refer to the highest temperature step at which a phase is part of the condensate assemblage in the 2K steps of the calculations. Our results are quite similar, but not identical, to those of Yoneda and Grossman (1995), referred to as the previous work in the following explanation of the differences which, in all cases, are due to differences in thermodynamic data. Note that, although some of the data used in our previous work may be more accurate, e.g., those for hibonite and grossite, we use those in Table 3 in the present study because the latter are more consistent with the MELTS liquid model. Hibonite forms from corundum 15K lower in the present calculations than in the previous work because hibonite is 2.5 kJ less stable and corundum 0.3 kJ more stable at 1700K in the present work. The lesser stability of hibonite in the present work allows it to be replaced by grossite and $CaAl_2O_4$, which are 16.7 and 8.7 kJ more stable, respectively, at 1700K in the present work. The gehlenite endmember of the melilite solid solution series is here 10.1 kJ less stable at 1600K than previously, and it forms from $CaAl_2O_4$, a phase more stable than hibonite in the present calculations. This causes the appearance temperature of melilite to be suppressed by nearly 50K and allows grossite and hibonite to partially replace it at lower temperature. Spinel condenses 13K lower in the present work than previously, primarily because the $MgAl_2O_4$ endmember is now 3.5 kJ less stable at 1500K. Plagioclase forms from spinel ~10K lower in the present work because the $CaAl_2Si_2O_8$ endmember is 6.5 kJ less stable at 1400K than previously. In the present work, $Ti_3O_5$ forms from Ti-bearing clinopyroxene 18K lower and $Ti_4O_7$ does not form at all because of gross differences in the way the Ti-bearing end-member components are treated in the two calculations. In the previous work, literature data were used for the $Ti^{3+}$-bearing component, $CaTiAlSiO_6$, and estimated for the $Ti^{4+}$-bearing component, $CaTiAl_2O_6$, while, in the present work, only data for the $Ti^{4+}$-bearing components, $CaTi_{0.5}Mg_{0.5}AlSiO_6$ and $CaTi_{0.5}Mg_{0.5}FeSiO_6$, are used. Cordierite replaces plagioclase in the present work because it is 12.8 kJ/mole more stable, and plagioclase is 7.2 kJ less stable, at 1300K relative to the previous work. Sphene does not form above 1200K in the present work because it is 5.5 kJ/mole less stable than previously. No liquids were found to be stable in solar gas at $P^{tot}$=$10^{-3}$ bar by us or any previous workers, e.g., Wagner (1979), Wood and Hashimoto (1993), and Yoneda and Grossman (1995), despite a contrary claim by Wark (1987).

**Table 6.** Temperatures (K) of appearance and disappearance of condensates from a gas of solar composition at $P^{tot} = 10^{-3}$ bar, compared with earlier results.

| | This work | | Yoneda and Grossman (1995) | |
|---|---|---|---|---|
| Mineral | In | Out | In | Out |
| Corundum | 1770 | 1726 | 1770 | 1740 |
| Hibonite | 1728 | 1686 | 1743 | 1500 |
| Grossite | 1698 | 1594 | | |
| Perovskite | 1680 | 1458 | 1688 | 1448 |
| $CaAl_2O_4$ | 1624 | 1568 | | |
| Melilite ss. | 1580 | 1434 | 1628 | 1444 |
| Grossite | 1568 | 1502 | | |
| Hibonite | 1502 | 1488 | | |
| Spinel ss. | 1488 | 1400 | 1501 | 1409 |
| Metal ss. | 1462 | | 1464 | |
| Clinopyroxene ss. | 1458 | | 1449 | |
| Olivine ss. | 1444 | | 1443 | |
| Plagioclase ss. | 1406 | 1318 | 1416 | |
| $Ti_3O_5$ | 1368 | 1342 | 1386 | 1361 |
| Orthopyroxene ss. | 1366 | | 1366 | |
| $Ti_4O_7$ | | | 1361 | 1217 |
| Cordierite | 1330 | | | |
| Cr-spinel ss. | 1230 | | 1221 | |
| Sphene | | | 1217 | |
| End of Computation | 1200 | | 970 | |

### 3.2. General Effects of Dust Enrichment and $P^{tot}$

Complete condensation calculations were performed from 2400K down to the last temperature step where our criteria for adequate convergence could be met, usually between 1100 and 1300K, and up to dust/gas enrichment factors of 1000x relative to solar composition, hereinafter abbreviated as "dust enrichments of 1000x". The extremes of $10^{-3}$ bar and $10^{-6}$ bar were chosen to bracket the generally accepted range of $P^{tot}$ in the inner solar nebula (Wood and Morfill, 1988). Results are shown at four different dust enrichments at $10^{-6}$ bar in Table 7 and at $10^{-3}$ bar in Table 8. In these tables, appearance and disappearance temperatures are defined as the highest temperature steps at which a phase is either part of or becomes absent from the stable condensate assemblage. In figures showing elemental distributions among coexisting phases, the fraction of an element present in a phase at its appearance temperature is extrapolated to zero in the next highest temperature step.

**Table 7.** Temperatures (K) of appearance and disappearance of condensates at $P^{tot} = 10^{-6}$ bar as a function of dust/gas enrichment.

| Dust/Gas Enrichment: | 1x | | 100x | | 500x | | 1000x | |
|---|---|---|---|---|---|---|---|---|
| Condensate | In | Out | In | Out | In | Out | In | Out |
| Corundum | 1570 | 1470 | 1890 | 1790 | 2010 | 1930 | 2040 | 1980 |
| Hibonite | 1480 | 1430 | 1790 | 1750 | 1940 | 1890 | 1980 | 1950 |
| Perovskite | 1460 | 1270 | 1690 | 1500 | 1760 | 1610 | 1740 | 1670 |
| Grossite | 1440 | 1370 | 1760 | 1650 | 1910 | 1780 | 1960 | 1820 |
| $CaAl_2O_4$ | 1390 | 1360 | 1680 | 1620 | 1820 | 1740 | | |
| Melilite | 1370 | 1250 | 1630 | 1470 | 1760 | 1740 | | |
| Grossite | 1360 | 1300 | 1620 | 1550 | 1740 | 1670 | | |
| Liquid | | | | | 1740 | 1730 | 1880 | 1370 |
| Melilite | | | | | 1730 | 1630 | | |
| Hibonite | 1320 | 1270 | 1550 | 1540 | | | | |
| Spinel | 1270 | 1210 | 1540 | 1430 | 1670 | 1610 | 1720 | 1660 |
| Liquid | | | | | 1630 | 1390 | | |
| Clinopyroxene | 1270 | | 1500 | | 1430 | | 1370 | |
| Olivine | 1240 | | 1490 | | 1610 | | 1660 | |
| Sapphirine | 1230 | 1190 | | | | | | |
| Metallic nickel-iron | 1210 | | 1360 | | 1420 | | 1430 | |
| Plagioclase | 1210 | 1190 | 1440 | 1300 | 1450 | | 1410 | |
| Cordierite | 1200 | | 1310 | | | | | |
| Orthopyroxene | 1190 | | 1400 | | 1490 | | 1530 | 1230 |
| Cr-spinel | 1160 | | 1410 | | 1610 | | 1660 | |
| Pyrophanite | | | 1130 | | 1230 | | 1250 | |
| MnO | | | 1090 | | 1180 | | 1210 | |
| End of Computation | 1100 | | 1090 | | 1160 | | 1200 | |

### *3.2.1. Oxygen Fugacity*

Shown in Fig. 4 is the temperature dependence of the oxygen fugacity of the gas in equilibrium with condensate assemblages computed at dust enrichments of 100x, 500x and 1000x at $P^{tot}=10^{-3}$ bar, and at dust enrichments of 100x and 1000x at $10^{-6}$ bar, along with that of the iron-wüstite buffer (log $f_{O_2}$ = IW) and that of a gas of solar composition at $10^{-3}$ bar (log $f_{O_2}$~IW-6) for reference. The curves for $P^{tot} = 10^{-3}$ bar are nearly concentric with one another and show the expected increase of $f_{O_2}$ with increasing dust enrichment. The curves for dust enrichments of 100x, 500x and 1000x lie at about IW-3.1, IW-1.7 and IW-1.2, respectively. Exceptions to this concentric behavior are seen as subtle changes in curvature, particularly noticeable at high dust enrichments where the onset of olivine condensation removes significant fractions of the oxygen from the vapor. Comparison of the two curves for a constant dust enrichment of 100x shows a slight increase in $fO_2$ by as much as 0.4 log units as $P^{tot}$ drops from $10^{-3}$ to $10^{-6}$ bar below 2000K. The smallness of the variation with $P^{tot}$ is due to the fact that $fO_2$ in oxygen-rich cosmic gases is largely controlled by the equilibrium $H_2 + {}^1/_2O_2 = H_2O$ and therefore depends on the $P_{H_2O}/P_{H_2}$ ratio which is almost independent of $P^{tot}$ at a given temperature, as discussed by Yoneda and Grossman (1995). Above 2000K, however, Fig. 4 shows that, at dust enrichments of 100x and 1000x, the $fO_2$ values at $10^{-6}$ bar drop below their respective values at $10^{-3}$ bar and the difference in $fO_2$ between the two total pressures at constant dust enrichment increases with increasing temperature, reaching nearly 4 log units at 2400K. This exceptionally large variation in $fO_2$with $P^{tot}$ is due to the fact that, at $10^{-6}$ bar, almost all of the $H_2$ and $H_2O$ are dissociated into monatomic species at these high temperatures, making the above equilibrium irrelevant to the $fO_2$while, at $10^{-3}$ bar, this dissociation occurs above 2400K because the higher pressure favors polyatomic over monatomic species.

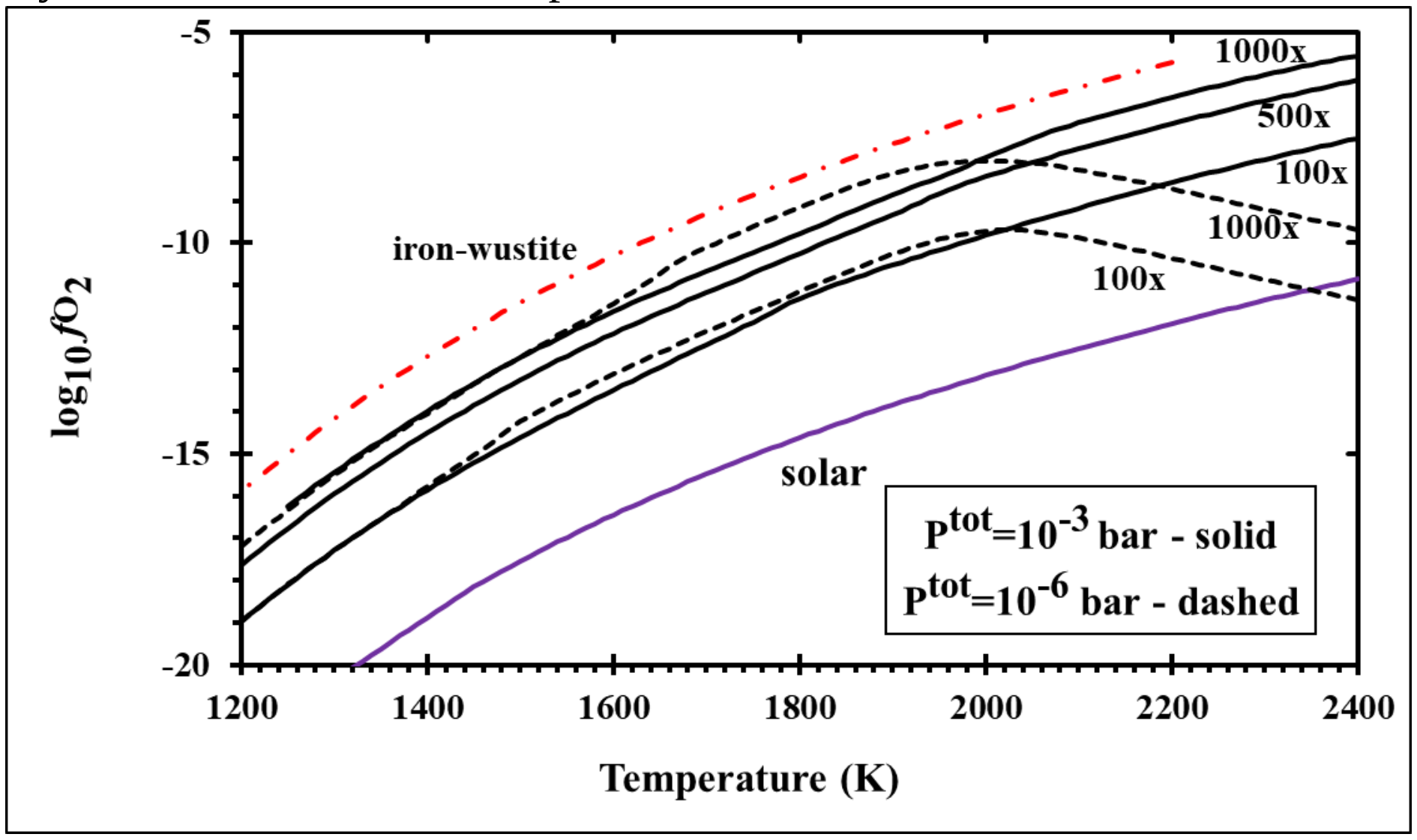

**Figure 4:** Variation of oxygen fugacity with temperature for gas in equilibrium with condensates at the stated conditions of total pressure and dust enrichment, with the iron-wüstite buffer curve (dash-dot red pattern) shown for reference.

### *3.2.2. Condensation Temperatures and Stability of Liquid*

The progressive increase in condensation temperatures of all phases with increasing dust enrichment at constant $P^{tot}$, as seen by Yoneda and Grossman (1995), is illustrated in Tables 7 and 8. At $10^{-6}$ bar, condensation temperatures are still low enough at a dust enrichment of 100x that no liquid phase is stable. At this $P^{tot}$ and above a dust enrichment between 400x and 450x, however, the oxide + silicate fraction of the assemblage that condenses in certain temperature intervals does so at a temperature above the solidus temperature for its bulk chemical composition, causing liquid to be a stable condensate. Upon cooling a system at $10^{-6}$ bar and a dust enrichment of 500x, liquid first appears at 1740K, where melilite and $CaAl_2O_4$ react with the gas to form grossite and a CMAS liquid. This liquid field persists for only 10K, at which point it crystallizes into melilite and grossite. A liquid field reappears at 1630K by reaction of melilite with the gas and persists to 1390K. At $10^{-6}$ bar and a dust enrichment of 1000x, condensation of all phases occurs at even higher temperatures such that a much greater range of bulk condensate compositions forms above solidus temperatures, causing the liquid stability field to extend to higher temperature, 1880K, and to persist without interruption to 1370K, replacing the stability fields of $CaAl_2O_4$, melilite and grossite.

**Table 8.** Temperatures (K) of appearance and disappearance of condensates at $P^{tot} = 10^{-3}$ bar as a function of dust/gas enrichment.

| Dust/Gas Enrichment: | 1x | | 100x | | 500x | | 1000x | |
|---|---|---|---|---|---|---|---|---|
| Condensate | In | Out | In | Out | In | Out | In | Out |
| Liquid | | | 2200 | 1390 | >2400 | 1400 | >2400 | 1310 |
| Corundum | 1770 | 1720 | | | | | | |
| Hibonite | 1720 | 1680 | | | | | | |
| Grossite | 1690 | 1590 | | | | | | |
| Perovskite | 1680 | 1450 | 1970 | 1810 | | | | |
| $CaAl_2O_4$ | 1620 | 1560 | | | | | | |
| Melilite | 1580 | 1430 | | | | | | |
| Grossite | 1560 | 1500 | | | | | | |
| Hibonite | 1500 | 1480 | | | | | | |
| Spinel | 1480 | 1400 | 1830 | 1710 | 1990 | 1940 | 2050 | 1990 |
| Metallic nickel-iron | 1460 | | 1690 | | 1780 | | 1800 | |
| Clinopyroxene | 1450 | | 1440 | | 1420 | | 1390 | |
| Olivine | 1440 | | 1780 | | 1940 | | 1990 | |
| Plagioclase | 1400 | 1310 | 1450 | | 1430 | | 1430 | |
| Orthopyroxene | 1360 | | 1620 | | 1700 | | | |
| ß-$Ti_3O_5$ | 1360 | 1340 | | | | | | |
| Cordierite | 1330 | | | | | | | |
| Cr-spinel | 1230 | | 1600 | | 1760 | | 1710 | |
| Pyrophanite | | | 1350 | | 1400 | | 1380 | |
| MnO | | | 1300 | | 1420 | | 1480 | |
| Pyrrhotite | | | | | 1330 | | 1380 | |
| Whitlockite | | | | | | | 1350 | |
| End of Computation | 1210 | | 1200 | | 1240 | | 1260 | |

At constant dust enrichment, condensation temperatures of all phases are higher at $10^{-3}$ bar than at $10^{-6}$ bar because partial pressures of most condensable elements increase with $P^{tot}$. As a result, the minimum dust enrichment necessary to condense partial melts at $10^{-3}$ bar is considerably lower than at $10^{-6}$ bar, and lies between 12x and 13x. At $10^{-3}$ bar, there is, at a dust enrichment of only 100x, an extensive and uninterrupted stability field of liquid extending up to 2200K and replacing the stability fields of corundum, hibonite, grossite, $CaAl_2O_4$ and melilite. At higher dust enrichments at this $P^{tot}$, the liquid stability field extends to even higher temperatures.

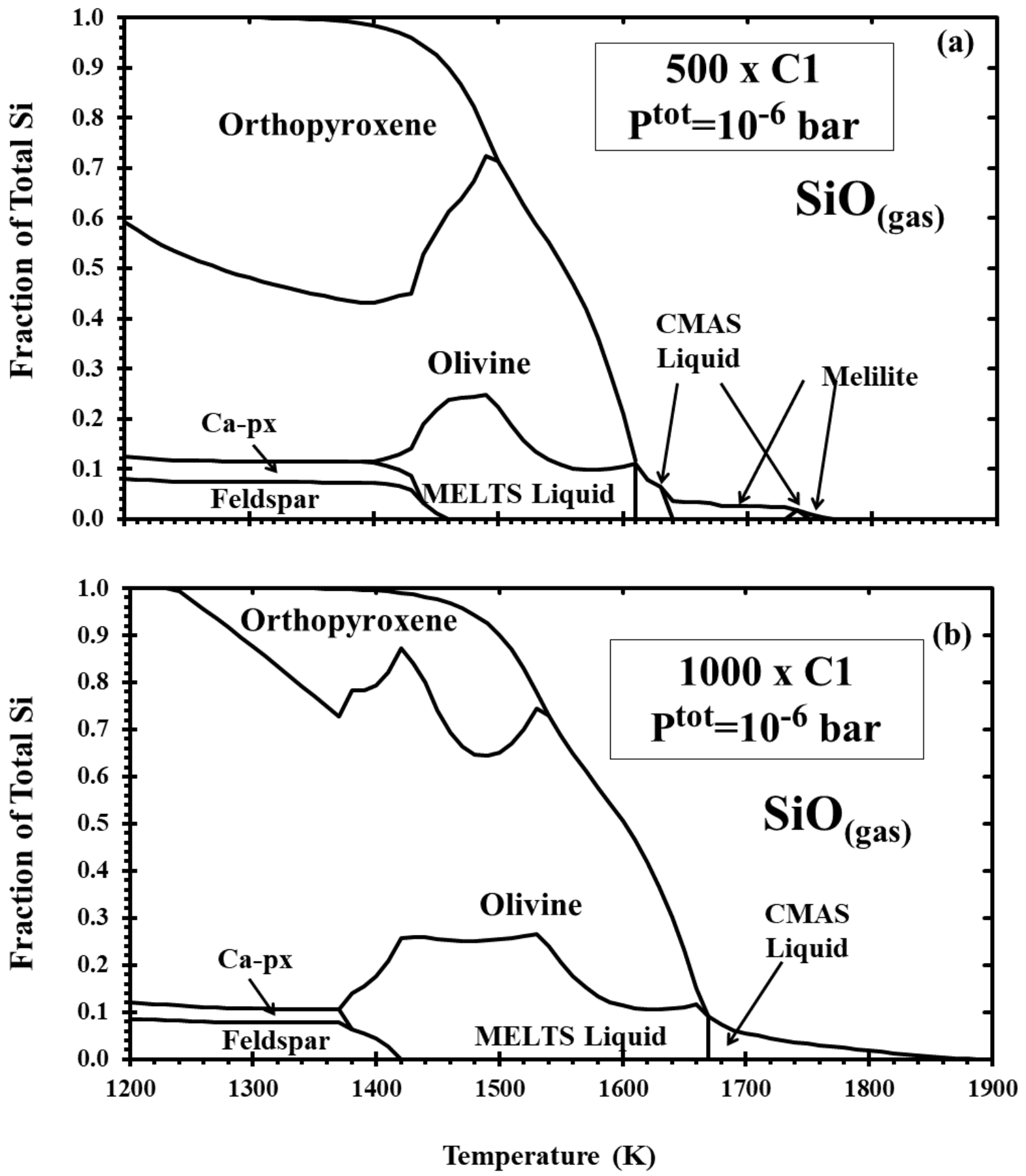

**Figure 5:** Distribution of Si between condensed phases and gas at $P^{tot} = 10^{-6}$ bar and a dust enrichment of (**a**) 500x; and (**b**) 1000x. Ca-px=Ca-rich clinopyroxene.

One way of viewing trends in the size of the liquid stability field as a function of $P^{tot}$ and dust enrichment is by comparison of graphs of the distribution of silicon between condensed phases and vapor *vs*. temperature. Such diagrams are presented in Figs. 5a and b for $10^{-6}$ bar and dust enrichments of 500x and 1000x, respectively, and should be compared to Figs. 6d and 7d for the cases of 100x and 1000x, respectively, at $10^{-3}$ bar. At $10^{-3}$ bar, the liquid has a stability field 800K wide (Table 8) and accounts for a maximum of 32% of the silicon at a dust enrichment of 100x. At the same $P^{tot}$, the liquid field widens to >1100K, with a maximum of 60% of the silicon, at a dust enrichment of 1000x. Although both the temperature interval for the stability of liquid and the maximum fraction of the total silicon accounted for by the liquid are always smaller at $10^{-6}$ bar than at $10^{-3}$ bar for the same dust enrichment, the liquid fields at $10^{-6}$ bar are still quite extensive at these elevated dust enrichments. For example, although a very small, high-temperature field of liquid is separated from a lower-temperature liquid field at 500x, the latter field is 240K wide and accounts for a maximum of 25% of the silicon and, at 1000x, the liquid field is over 500K wide and accounts for a maximum of 26% of the silicon. Petaev and Wood (1998) found no liquid stability field at $10^{-5}$ bar and a dust enrichment of $10^7$. This is in clear disagreement with the results of Wood and Hashimoto (1993) who found a small liquid stability field at the same $P^{tot}$ and a lower dust enrichment, $10^3$, as both studies employed the same liquid solution model and dust composition. The Petaev and Wood (1998) results are also in complete disagreement with ours in that we find a liquid stability field at both lower $P^{tot}$ and lower dust enrichment. Although the existence of a liquid stability field in Wood and Hashimoto's (1993) study is in general agreement with the work presented here, both the amount of liquid and the temperature interval of its stability are much smaller than would be expected from our work, presumably because of Wood and Hashimoto's (1993) use of an ideal solution model for silicate liquids, which are demonstrably non-ideal (Berman, 1983; Ghiorso *et al.*, 1983), and the difference between their assumed dust composition and ours.

From Tables 7 and 8, it is clear that the assemblage liquid + metallic nickel-iron + olivine + orthopyroxene + Cr-spinel occupies a very wide stability field within the ranges of $P^{tot}$ and dust enrichment considered herein.

### 3.3. Condensation at 100x dust enrichment and $P^{tot}=10^{-3}$ bar

The distributions of Al, Mg, Ca, Si, Fe, and Na and K between condensed phases and vapor are illustrated in Figs. 6a, b, c, d, e, and f, respectively, for a system enriched 100x in dust at $P^{tot}=10^{-3}$ bar. The first condensate, a CMAS liquid extremely rich in $Al_2O_3$ (79 mole %) and CaO (21%) with only minor amounts of MgO and $SiO_2$, condenses at 2200K. Perovskite becomes stable at 1970K and consumes 85% of the total Ti in the system by 1850K. At 1830K, gaseous Mg, Fe and Cr begin to react with the liquid and perovskite to form an Mg-, Al-rich spinel with minor amounts of Cr, Fe and Ti. By 1810K, this spinel has molar Fe/Mg and Cr/Al ratios of $3.4\text{x}10^{-3}$ and $1.3\text{x}10^{-2}$, respectively, and a $TiO_2$ content of 9.1 wt%, so much Ti that perovskite disappears at this temperature. With falling temperature, Ti and Mg continue to condense into spinel, and Si and Mg into the liquid. The $SiO_2$ and MgO contents of the liquid increase, and the $MgAl_2O_4$ component of the spinel dissolves into the liquid. By 1780K, the liquid reaches ~40 wt% $SiO_2$ and ~26% MgO, olivine (0.25 wt% FeO, 0.73% CaO) becomes stable and, as discussed above, this triggers the switchover from the CMAS to the MELTS liquid model. At this point, the spinel has molar Fe/Mg and Cr/Al ratios of $3.0x10^{-3}$ and $2.3x10^{-2}$, respectively, and, because Ti can now be accommodated by the liquid model, the $TiO_2$ content of the liquid is 0.8 wt% and Ti drops to only 0.44 wt% in the spinel. With falling temperature, most olivine forms by wholesale condensation of Mg and Si from the gas but some by crystallization of the liquid. With the total amount of liquid decreasing slightly, the $MgAl_2O_4$ component of spinel continues to dissolve into it, causing the molar Fe/Mg and Cr/Al ratios in spinel to rise to $9.0x10^{-3}$ and $6.2x10^{-2}$, respectively, just before spinel dissolves completely into the liquid at 1710K. With falling temperature, the FeO content of olivine increases and its CaO decreases, reaching 0.89 wt% and 0.48%, respectively, at the point where spinel disappears and 1.29% and 0.42%, respectively, at 1690K, the initial condensation temperature of metallic NiFe. This alloy contains 13.6 wt% Ni, 0.48% Co and 0.24% Cr. As the temperature falls, olivine of increasing FeO and decreasing CaO content continues to condense from the gas, Si and Fe continue to condense into the liquid, diluting its MgO, $Al_2O_3$ and CaO contents, and metal of decreasing Ni and Co and increasing Cr content continues to condense. At 1620K, when nearly all the Mg and 80% of the Si are condensed, gaseous SiO begins to react with olivine and liquid to form orthopyroxene with an initial FeO content of 1.1 wt%. At this point, olivine contains 1.9 wt% FeO and 0.23% CaO and the liquid contains 0.59 wt% $TiO_2$, 0.23% $Cr_2O_3$ and 0.30% FeO. At 1600K, gaseous Cr begins to react with the liquid to form a small amount of Cr-spinel, having molar Fe/Mg and Cr/Al ratios of 0.049 and 2.1, respectively, and a $TiO_2$ content of 0.26 wt%. With falling temperature, orthopyroxene of increasing FeO content continues to form by reaction of gaseous SiO with liquid and with olivine of increasing FeO and CaO contents; all components of the metal alloy continue to condense, resulting in decreasing Ni, Co and Cr contents; the amount of Cr-spinel continues to increase at the expense of $Al_2O_3$ in the liquid and gaseous Cr; and the CaO and $TiO_2$ contents of the liquid increase while its FeO and $Cr_2O_3$ contents decrease. By 1550K, 1.8% of the K has condensed into the liquid which contains 0.01wt% $K_2O$. At 1450K, 99% of the Fe is condensed, and gaseous Na begins to react with the liquid to form plagioclase feldspar containing 1.6 mole % albite. At this point, olivine contains 2.30 wt% FeO and 0.34% CaO, Cr-spinel has molar Fe/Mg and Cr/Al ratios of 0.053 and 1.43, respectively, and 0.34 wt% $TiO_2$, and liquid contains 0.93 wt% $TiO_2$, 0.20% $Cr_2O_3$, 0.19% FeO, 0.06% $K_2O$ and 0.01% $Na_2O$. At 1440K, a diopsidic clinopyroxene begins to co-crystallize with feldspar from the remaining liquid, which reaches 0.15 wt% $K_2O$ and 0.03% $Na_2O$ before disappearing at 1390K.

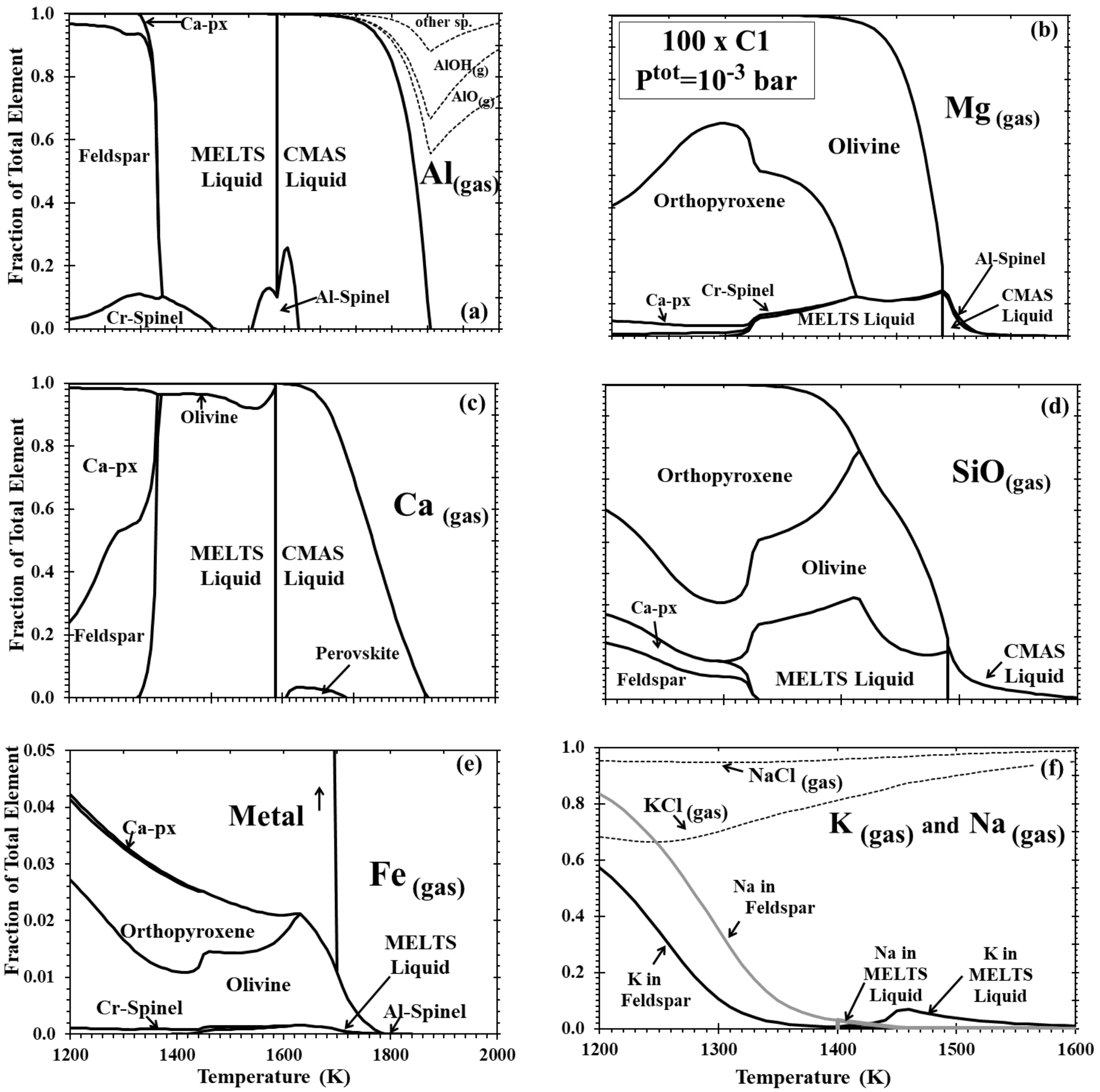

**Figure 6:** Distribution of (a) Al; (b) Mg; (c) Ca; (d) Si; (e) Fe; and (f) Na and K between condensed phases and vapor at a dust enrichment of 100x at $P^{tot} = 10^{-3}$ bar. Note vertical scale change in (e). Al-spinel=spinel with ≤ 1 wt% $Cr_2O_3$; Cr-spinel=spinel with >1 wt% $Cr_2O_3$. Other abbreviations as used previously.

With continued cooling, gaseous Na and K react with anorthitic feldspar, increasing its albite and orthoclase contents and displacing Ca which, in turn, reacts with orthopyroxene to form clinopyroxene and olivine. The FeO contents of olivine, orthopyroxene, clinopyroxene and spinel all increase as metallic Fe becomes oxidized, causing the Ni and Co contents of the alloy to increase even as its Cr content decreases due to reaction with spinel. At 1350K, gaseous Mn reacts with Ti in the spinel and clinopyroxene to form an oxide solid solution consisting of pyrophanite [$MnTiO_3$] with 9.4 wt% MgO and 2.6% FeO. At 1300K, the remaining gaseous Mn begins to condense as MnO. By 1200K, olivine contains 3.6 wt% FeO and 0.10% CaO; clinopyroxene contains 1.4% $Al_2O_3$, 0.62% FeO, 0.25% $TiO_2$, and 0.27% $Na_2O$; spinel has molar Fe/Mg and Cr/Al ratios of 0.13 and 5.1, respectively, and a $TiO_2$ content of 0.59%; and the oxide solid solution contains only 1.3% MgO and 1.6% FeO. At this temperature, Na, K and Mn are only partially condensed, with 15, 43 and 3.5%, respectively, remaining in the vapor.

### 3.4. Condensation at 1000x and $P^{tot}=10^{-3}$ bar

The distributions of Al, Mg, Ca, Si, Fe, and Na and K between condensed phases and vapor are illustrated in Figs. 7a, b, c, d, e, and f, respectively, for a system enriched 1000x in dust at $P^{tot}=10^{-3}$ bar. At a dust enrichment of 1000x, a CMAS liquid is already present at 2400K, into which 75% of the Al and 27% of the Ca have condensed. At 2050K, gaseous Ti, Mg, Cr and Fe begin to react with $Al_2O_3$ in the liquid to form a spinel containing 44.8 wt% MgO, 39.5% $TiO_2$, 12.7% $Al_2O_3$, 1.8% $Cr_2O_3$, 1.1% FeO and 0.12% $Fe_2O_3$. At 1990K, olivine begins to form primarily by condensation from the gas but some also by crystallization from the liquid, and the switch is made from the CMAS to the MELTS liquid model which, because the latter can accommodate $TiO_2$, causes the titanian spinel to dissolve into the liquid. The initial olivine contains 0.92 wt% FeO and 0.23% CaO but, as Fe and Mg condense into it, reaches 10.8% FeO and 0.13% CaO by 1800K. Over the same temperature range, as Fe and Cr condense into the liquid, the composition of the latter evolves from 0.73 wt% FeO, 0.39% $TiO_2$, 0.02% $Cr_2O_3$ and <0.01% $Fe_2O_3$ to 24.2% FeO, 0.23% $TiO_2$, 0.90% $Cr_2O_3$ and 0.19% $Fe_2O_3$. Metal alloy containing 20.0 wt% Ni, 0.70% Co and 0.09% Cr begins to condense at 1800K, and reaches 11.7% Ni, 0.52% Co and 0.06% Cr by 1720K. In this temperature range, olivine continues to form at the expense of liquid and reaches 14.3 wt% FeO and 0.13% CaO, while the liquid composition evolves to 25.9 wt% FeO, 0.25% $TiO_2$, 1.22% $Cr_2O_3$ and 0.16% $Fe_2O_3$. At 1710K, a small amount of Cr-spinel begins to form by drawing Al and most of its Cr from the liquid but some of its Cr also from the gas and the metal. Initially, it has molar Fe/Mg and Cr/Al ratios of 0.62 and 3.79, respectively, and contains 1.10 wt% $Fe_2O_3$ and 0.27% $TiO_2$ but varies in composition as it continues to crystallize from the liquid with falling temperature, reaching molar Fe/Mg and Cr/Al ratios of 1.34 and 2.29, respectively, with 1.42 wt% $Fe_2O_3$ and 1.07% $TiO_2$ at 1440K. Over the same temperature range, the amount of olivine continues to increase with falling temperature, drawing its MgO, $SiO_2$ and CaO from the liquid. From 1710 to 1560K, the FeO consumed by olivine comes from both liquid and gas but, at 1560K, the temperature below which <1% of the Fe remains in the gas, oxidation of the metal alloy joins the liquid as a source of the FeO for continued production of olivine. The amount of metal alloy increases with falling temperature from 1710 to 1560K, as gaseous Fe continues to condense into it, diluting its Ni, Co and Cr concentrations to 9.3% Ni, 0.43% Co and 0.01% Cr at 1560K. Below 1560K, however, oxidation of Fe causes the amount of metal to decrease with falling temperature, increasing its Ni and Co contents to 10.0 and 0.46 wt%, respectively, by 1440K. Its Cr content continues to decrease due to formation of increasing amounts of Cr-spinel with falling temperature. Olivine contains 21.1 wt% FeO and 0.24% CaO at 1560K, and 24.6% FeO and 0.45% CaO at 1440K. As the amount of liquid decreases with falling temperature, its FeO, $Cr_2O_3$ and $Fe_2O_3$ contents progressively decrease, reaching 17.6 wt%, 0.45% and 0.02%, respectively, at 1560K and 10.8%, 0.12% and 0.01% at 1440K; and its $TiO_2$, $Na_2O$, $K_2O$ and $P_2O_5$ contents progressively increase, reaching 0.43 wt%, 0.29%, 0.10% and <0.01%, respectively, at 1560K and 0.56%, 2.79%, 0.35% and 0.20% at 1440K. At 1480K, MnO condenses and, at 1430K, plagioclase containing 34.4 mole % albite and 0.35% orthoclase begins to crystallize from the liquid. As the amount of plagioclase increases with falling temperature, its albite and orthoclase contents also increase. Although the Na required for this is supplied by both gas and liquid, the K is derived only from the liquid, with the proportion of K residing in the vapor actually increasing initially with falling temperature. At 1390K, a diopside-rich clinopyroxene, containing 4.7 wt% FeO, 2.3% $Al_2O_3$, 0.18% $Na_2O$, 0.36% $TiO_2$ and 0.09% $Fe_2O_3$, crystallizes from the liquid. At 1380K, a pyrophanite-rich oxide solid solution, containing 9.53 wt% FeO, 1.02% MgO and 0.52% $Fe_2O_3$, forms by reaction of gaseous Mn with $TiO_2$ in the liquid; and gaseous sulfur begins to react with metallic Fe to form pyrrhotite, $Fe_{0.877}S$. The concentrations of Ni and Co in the residual alloy are 11.5 wt% and 0.53%, respectively, but increase sharply as more pyrrhotite forms with falling temperature, reaching 21.4% and 0.98%, respectively, at 1310K. At 1350K, gaseous P reacts with the liquid to form whitlockite. At 1320K, just before disappearing, the liquid contains 10.1 wt% $Na_2O$, 4.52% FeO, 1.34% $K_2O$, 0.77% $P_2O_5$, 0.41% $TiO_2$ and 0.02% $Cr_2O_3$. At 1260K, olivine contains 27.3 wt% FeO and 0.31% CaO; clinopyroxene contains 4.5 wt% FeO, 1.6% $Al_2O_3$, 0.33% $Na_2O$, 0.33% $TiO_2$, and 0.07% $Fe_2O_3$; Cr-spinel has molar Fe/Mg and Cr/Al ratios of 2.9 and 7.5, respectively, and contains 2.8 wt% $TiO_2$ and 1.5% $Fe_2O_3$; the pyrophanite-rich solid solution contains 9.2 wt% FeO, 0.86% MgO and 0.45% $Fe_2O_3$; and the metal alloy contains 27.5 wt% Ni and 1.3% Co. At this point, 97.8% of the P is condensed as whitlockite, 65.1% of the sulfur as pyrrhotite, and 90.0% of the Na and 65.5% of the K as feldspar. The ratio of the proportion of Fe in sulfide to that in metal is 2.2.

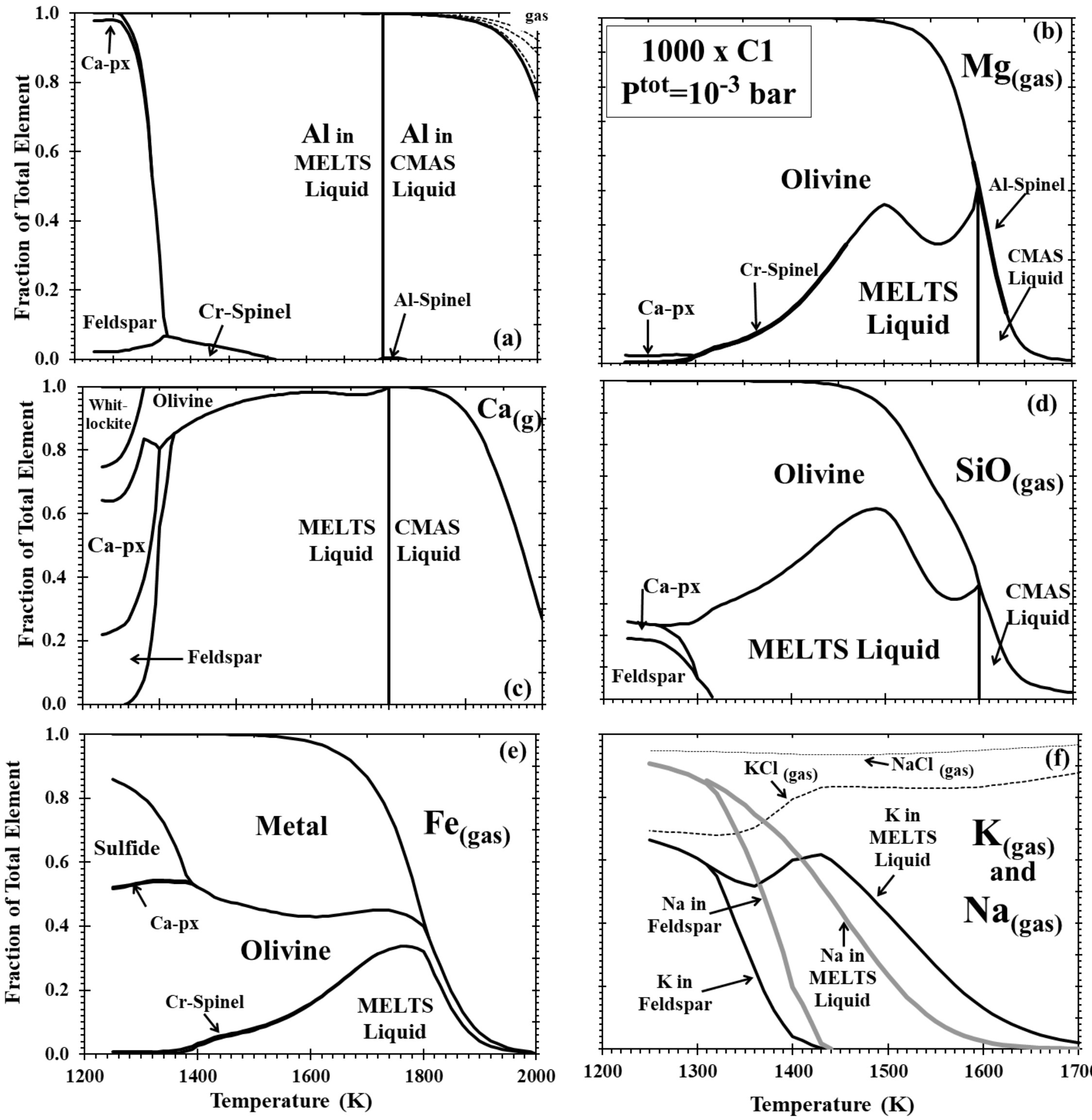

**Figure 7:** Distribution of (a) Al; (b) Mg; (c) Ca; (d) Si; (e) Fe; and (f) Na and K between condensed phases and vapor at a dust enrichment of 1000x at $P^{tot} = 10^{-3}$ bar. Abbreviations as used previously.

### 3.5. Direct Condensation of Oxidized Iron at High Temperature

The historical motivation behind studying dust enrichment in the solar nebula was to increase the oxygen fugacity so as to produce condensates that are more oxidized than is possible in a gas of solar composition. For example, one of the most perplexing problems for condensation theory is how to produce the observed molar Fe/Fe+Mg ratios that range from 0.2 in olivine and pyroxene in H-Group ordinary chondrites, to 0.3 in the LL-Group, to 0.5 or above in the matrices of some CV3 chondrites. Solar gas is so reducing that all Fe condenses initially as metal. In order to produce FeO-rich silicates, oxidation of this metal, which occurs only at or below ~500K, must be followed by diffusion of $Fe^{2+}$ through the crystal structures of pre-existing silicates. This requires equilibration between solids and a low-density gas and efficient diffusion through nearly close-packed silicate structures at temperatures where the rates of these processes are so low that most workers doubt chondritic matter obtained its oxidation state in this way.

The extent to which this problem is alleviated here is illustrated by the fayalite content of olivine and ferrosilite content of orthopyroxene, which are plotted as functions of temperature at various dust enrichments at $10^{-3}$ bar in Figs. 8a and b and at $10^{-6}$ bar in Figs. 8c and d. In the dust-enriched systems considered in these calculations, the oxygen fugacity is so high that significant amounts of FeO are stable in olivine and pyroxene at very high temperatures, where gas-solid equilibration is much more likely and diffusion rates are expected to be much higher than at the low temperatures where FeO would become stable in a solar gas. In dust-enriched systems, not only does the first-condensing olivine contain significant quantities of FeO, but also, as the temperature falls below the initial condensation temperature of olivine, the equilibrium fayalite content is predicted to increase while additional olivine condenses from the gas. In this temperature interval of direct olivine condensation, $Fe^{2+}$ is incorporated into each olivine grain as it grows directly from the vapor phase. The assumption of thermodynamic equilibrium requires that all olivine at a given temperature have the same fayalite content, and that is the value computed here. This requires that the relatively fayalite-poor olivine condensed at high temperature must become as fayalite-rich as newly condensing olivine at a lower temperature. If, even at the high temperatures being discussed here, the $Fe^{2+}$ contents of the interiors of the previously formed olivine crystals cannot increase fast enough to maintain equilibrium, the newly condensing olivine at any temperature will have even higher fayalite contents than calculated here. For all dust enrichments considered here at $10^{-6}$ bar, and for dust enrichments <800x at $10^{-3}$ bar, we can consider this stage, olivine formation by direct condensation from the gas, to end at the temperature where olivine begins to react with the gas to form orthopyroxene. For dust enrichments $\geq$ 800x at $10^{-3}$ bar, however, direct condensation of olivine ceases at a higher temperature than that for orthopyroxene formation because the fraction of the total Si remaining in the gas becomes very small, <1%. The trajectory of the temperature of cessation of direct olivine condensation is marked on Figs. 8a and 8c. At both values of $P^{tot}$, a liquid is present which persists to temperatures well below that where orthopyroxene begins forming. The FeO contents of olivine and orthopyroxene continue to rise with falling temperature and, with liquid present, this can occur by equilibration of olivine and orthopyroxene with the liquid. This is a much more kinetically favorable process for forming FeO-rich silicates than equilibration with a low-density gas.

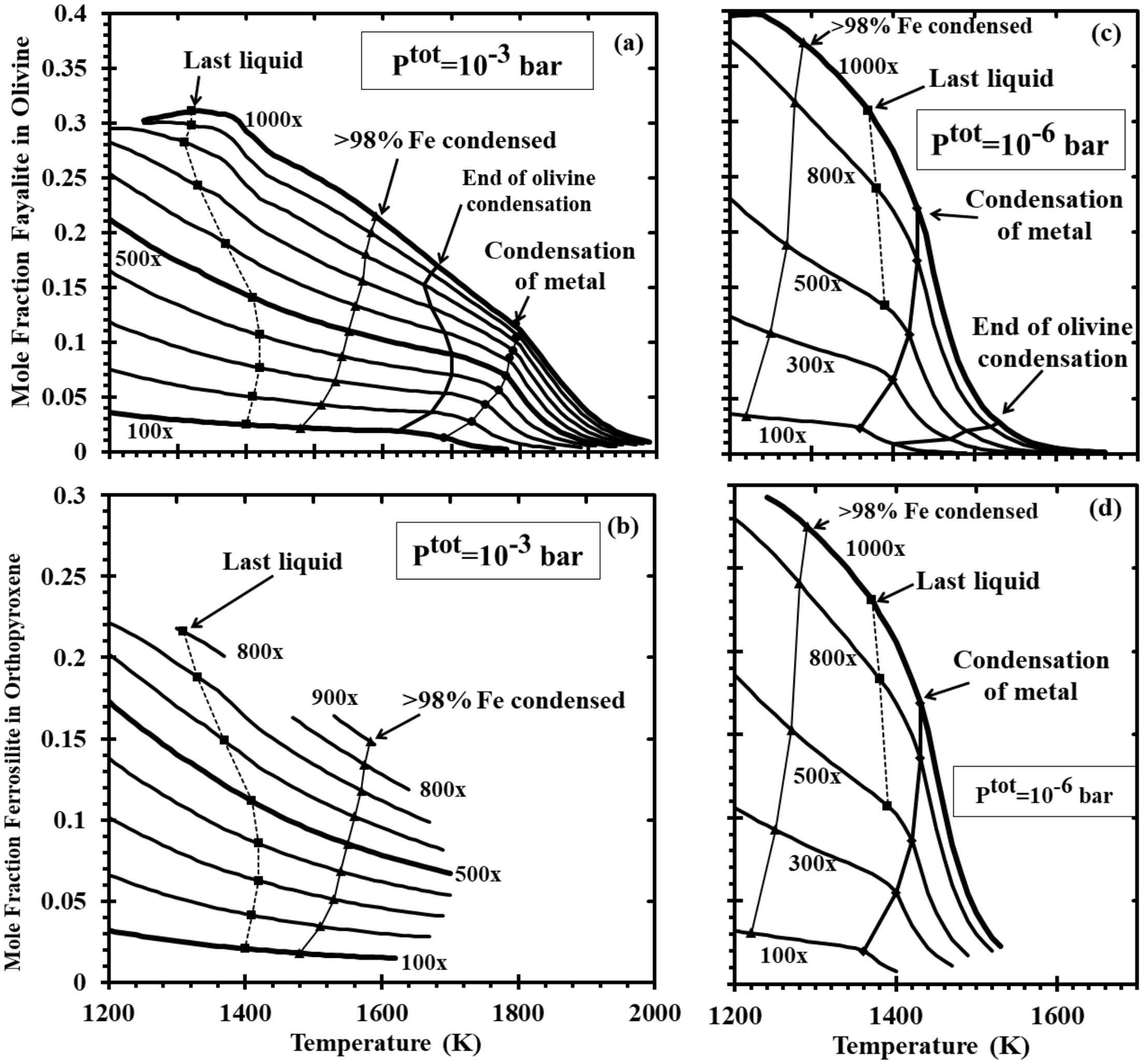

**Figure 8:** Mole fraction of iron endmember at $10^{-3}$ bar in (a) olivine and (b) orthopyroxene and at $10^{-6}$ bar in (c) olivine and (d) orthopyroxene as a function of temperature and at dust enrichment factors with integral multiples of 100. Trajectories of the condensation temperature of metallic nickel-iron alloy, the temperature at which direct condensation of olivine ceases, the temperature at which >98% of the total iron is condensed, and the temperature of disappearance of silicate liquid are indicated.

At $10^{-3}$ bar, $X_{Fa}$ for the first-condensing olivine is only $2.5 \times 10^{-3}$, $5.1 \times 10^{-3}$ and $9.1 \times 10^{-3}$ at dust enrichments of 100x, 500x and 1000x, respectively. At the temperature where olivine condensation ends, however, $X_{Fa}$ has increased to 0.019, 0.088 and 0.164 at dust enrichments of 100x, 500x and 1000x, respectively, and, at the temperature of disappearance of liquid, $X_{Fa}$ is 0.025, 0.14 and 0.31. At $10^{-6}$ bar, $X_{Fa}$ is lower in the first-condensing and last-condensing olivine than at $10^{-3}$ bar and the same dust enrichment, and almost the same at the temperature where the liquid disappears. Below the solidus, continued increase in the FeO content of olivine with decreasing temperature is governed by multiphase equilibrium, which would be impeded not only by slow diffusion of $Fe^{2+}$ into silicates but also by slow solid-state reaction rates between olivine, orthopyroxene, clinopyroxene and plagioclase, and by slow reaction between gaseous oxidizing agents and metallic iron. If the arbitrary assumption is made that multiphase equilibrium can be maintained to temperatures as low as 1200K, nebular olivine can be expected to have $X_{Fa}$ of 0.036 and 0.21 at dust enrichments of 100x and 500x, respectively, at $10^{-3}$ bar, and 0.036, 0.23 and 0.40 at dust enrichments of 100x, 500x and 1000x, respectively, at $10^{-6}$ bar. Higher FeO contents require equilibration to lower temperatures under these conditions or condensation at higher dust enrichments.

At given $P^{tot}$, the temperature at which gaseous SiO begins to react with olivine to form orthopyroxene increases, reaches a maximum and finally decreases with progressively increasing dust enrichment. At a dust enrichment of 1000x, there is no orthopyroxene stability field at all at $10^{-3}$ bar. At relatively low dust enrichments, the orthopyroxene condensation temperature increases with increasing dust enrichment due to the attendant increase in the partial pressure of SiO. Accompanying this, however, is an increase in the (FeO+MgO)/$SiO_2$ ratio of the silicate fraction of the condensate due to the increase in oxygen fugacity with increasing dust enrichment. This tends to stabilize olivine at the expense of orthopyroxene. A dust/gas ratio is reached beyond which the stabilizing effect on orthopyroxene of the increasing partial pressure of SiO is outweighed by the stabilizing effect on olivine of the increasing (FeO+MgO)/$SiO_2$ ratio. This causes the orthopyroxene field to shrink at the expense of the olivine field with increasing dust/gas ratio, and eventually disappear altogether.

### 3.6. Bulk Chemical Composition of Condensates

In Fig. 9a and b, the bulk chemical composition of the total condensate is plotted as a function of temperature at a $P^{tot}$ of $10^{-3}$ bar and a dust enrichment of 560x. Features of this diagram which are common to condensation at all dust enrichments are the early entry of Al, Ca and Ti relative to Mg and Si, as well as the relatively late entry of Na, K and Mn into the condensates. Features specific to condensation at this dust enrichment are the relative proportions of metal, FeO and sulfide as a function of temperature. This particular dust enrichment was chosen because it yields a single temperature at which the distribution of Fe between metal, sulfide and silicate matches closely the distribution found in H-Group ordinary chondrites and results in a bulk chemical composition very close to the average of those meteorites.

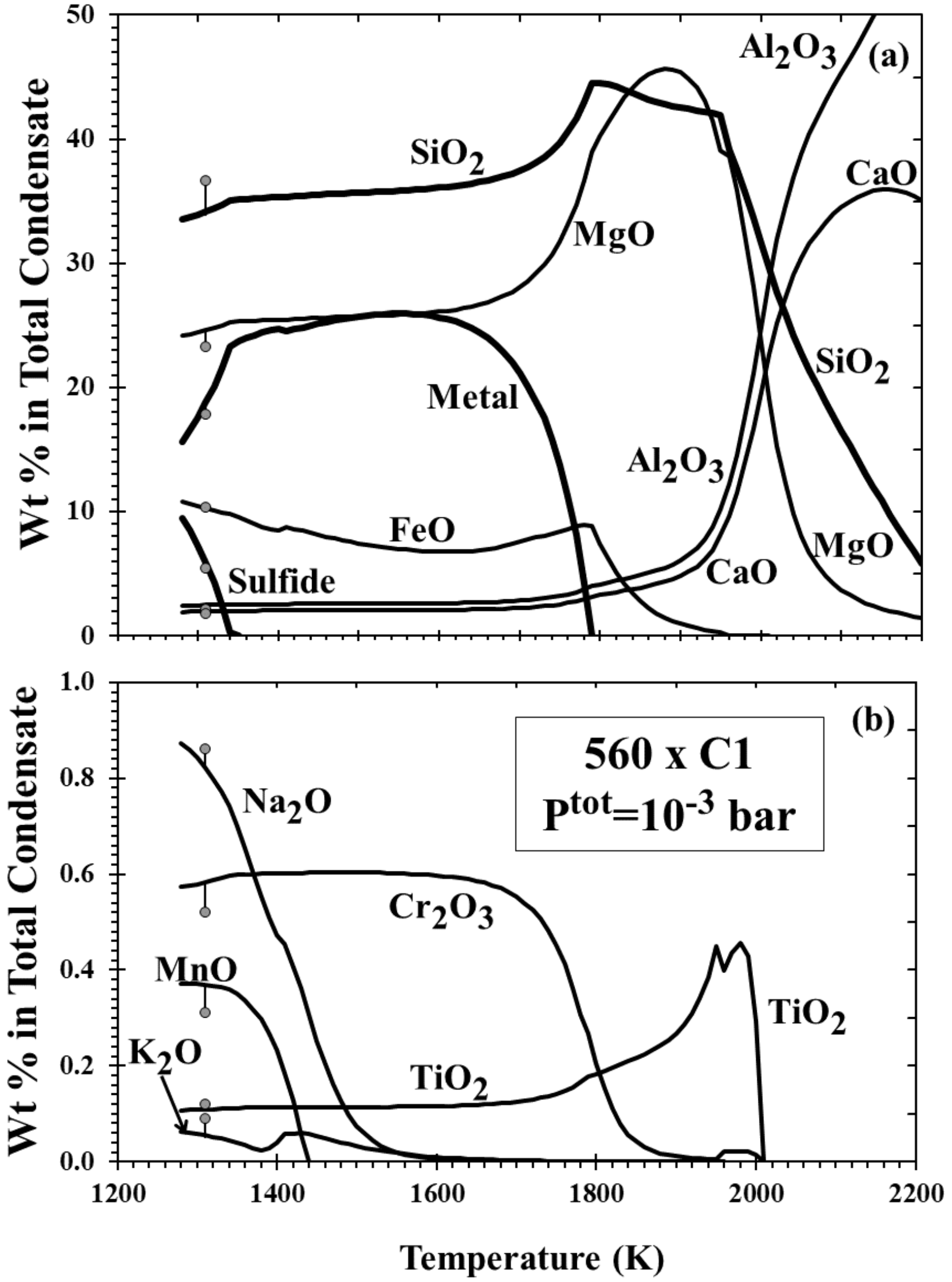

**Figure 9:** Concentrations of (a) major and (b) minor components of the total condensate as a function of temperature at the stated conditions. Filled circles indicate average composition of H-Group chondrites.

For example, at 1310K, the total condensate and, for comparison (brackets), the average H-Group chondrite fall from Jarosewich (1990) contains 18.8 (17.8) wt % Fe+Ni+Co metal, 10.2 (10.3) % FeO, 5.9 (5.4) % FeS, 34.2 (36.6) % $SiO_2$ and 24.6 (23.3) % MgO. Similarly, at the same $P^{tot}$ and a slightly higher dust enrichment of 675x, a temperature can be found at which the bulk chemical composition of the condensate comes very close to the average composition of L-Group chondrite falls from Jarosewich (1990). In the following comparison, sufficient metal of the same composition as that in Jarosewich's average L-Group chondrite has been added to his average L-Group chondrite bulk composition to yield the same atomic Fe/Si ratio as in H-Group chondrites. At 1330K, the condensate contains 16.8 (16.1) wt % metal, 12.9 (13.2) % FeO, 5.5 (5.3) % FeS, 34.0 (36.3) % $SiO_2$ and 24.5 (22.6) % MgO. In both cases, the $MgO/SiO_2$ ratio of the condensate is higher than in the chondrite due to the fact that the relative abundances of non-volatile elements in the model system are those of C1 chondrites, which are known to have a higher atomic Mg/Si ratio than ordinary chondrites. Nevertheless, the close correspondence in composition between the predicted condensates and the chondrite averages serves to emphasize the point that the distribution of iron between metal, silicate and sulfide in ordinary chondrites could have been established during high-temperature condensation in a dust-enriched system.

### 3.7. Composition of Silicate Liquid

The temperature variation of the composition of the silicate melt is shown for the cases of $10^{-3}$ bar and a dust enrichment of 100x in Figs. 10a and b, $10^{-3}$ bar and a dust enrichment of 1000x in Figs. 10c and d, and for $10^{-6}$ bar and a dust enrichment of 1000x in Figs.10e and f. The evolution of the liquid composition is similar in all cases, but some exceptions are noteworthy. Because Al is more refractory than Ca, the $Al_2O_3$ content of the initial liquid is very high but falls with decreasing temperature due to dilution by CaO which condenses more gradually with falling temperature. Similarly, at lower temperatures, incipient condensation of more volatile Si and Mg into the liquid causes the concentrations of $SiO_2$ and MgO to increase, diluting both $Al_2O_3$ and CaO. Comparing Figs. 10a and c, it is seen that, for liquids that form at higher temperatures than olivine, similar liquid compositions are stable at temperatures 300K higher when the dust enrichment is increased by a factor of 10 at $10^{-3}$ bar. Similarly, comparing Figs. 10c and e reveals that liquids of similar composition form about 400K higher when $P^{tot}$ is increased by a factor of 1000 at a dust enrichment of 1000x. Note, however, that, at the condensation temperatures of olivine, liquid compositions are quite different from one another at different combinations of $P^{tot}$ and dust enrichment. For example, the concentrations of MgO and $SiO_2$ in the liquid at a dust enrichment of 1000x are almost the same where olivine condenses at $10^{-3}$ bar (Fig. 10c) but the $SiO_2$ content is more than double that of MgO at $10^{-6}$ bar (Fig. 10e). One way of understanding this is by considering the fact that the solubility of olivine in a melt of a given composition is quite different at temperatures hundreds of degrees apart.

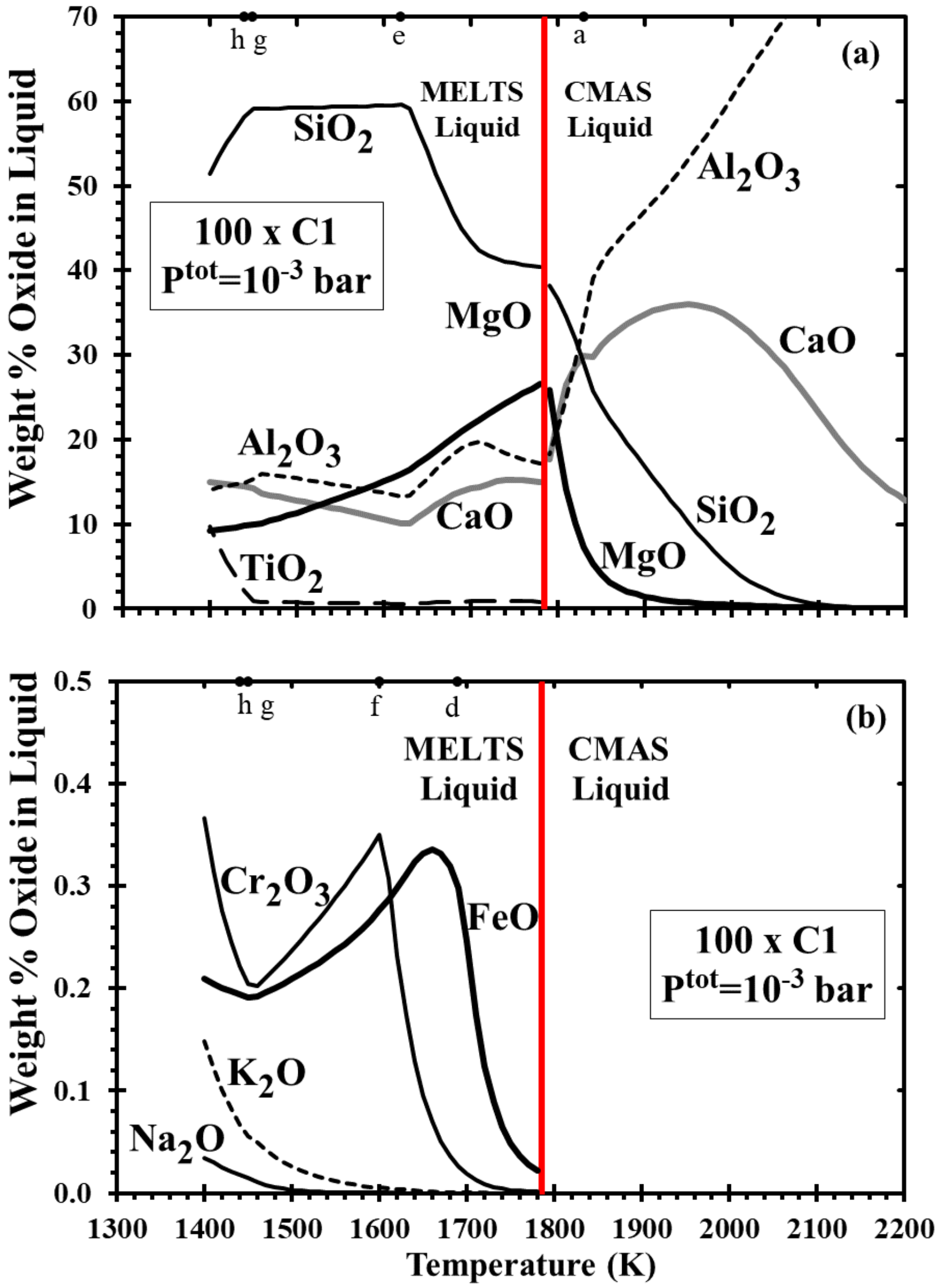

**Figure 10:** Compositions of condensate liquids at (a), (b) $10^{-3}$ bar and a dust enrichment of 100x; (c), (d) $10^{-3}$ bar and a dust enrichment of 1000x; and (e), (f) $10^{-6}$ bar and a dust enrichment of 1000x. In all cases, the vertical line marks the condensation temperature of olivine, where the transition between CMAS and MELTS liquid models is made. In (e) and (f), $Na_2O \leq 0.01$ wt%. Inflection points are due to the onset of crystallization or disappearance of a coexisting phase, and are labelled as follows: a, spinel in; d, metal in; e, orthopyroxene in; f, Cr-spinel in; g, feldspar in; h, clinopyroxene in; k, rhombohedral oxide in; p, perovskite in; q, clinopyroxene in and liquid out.

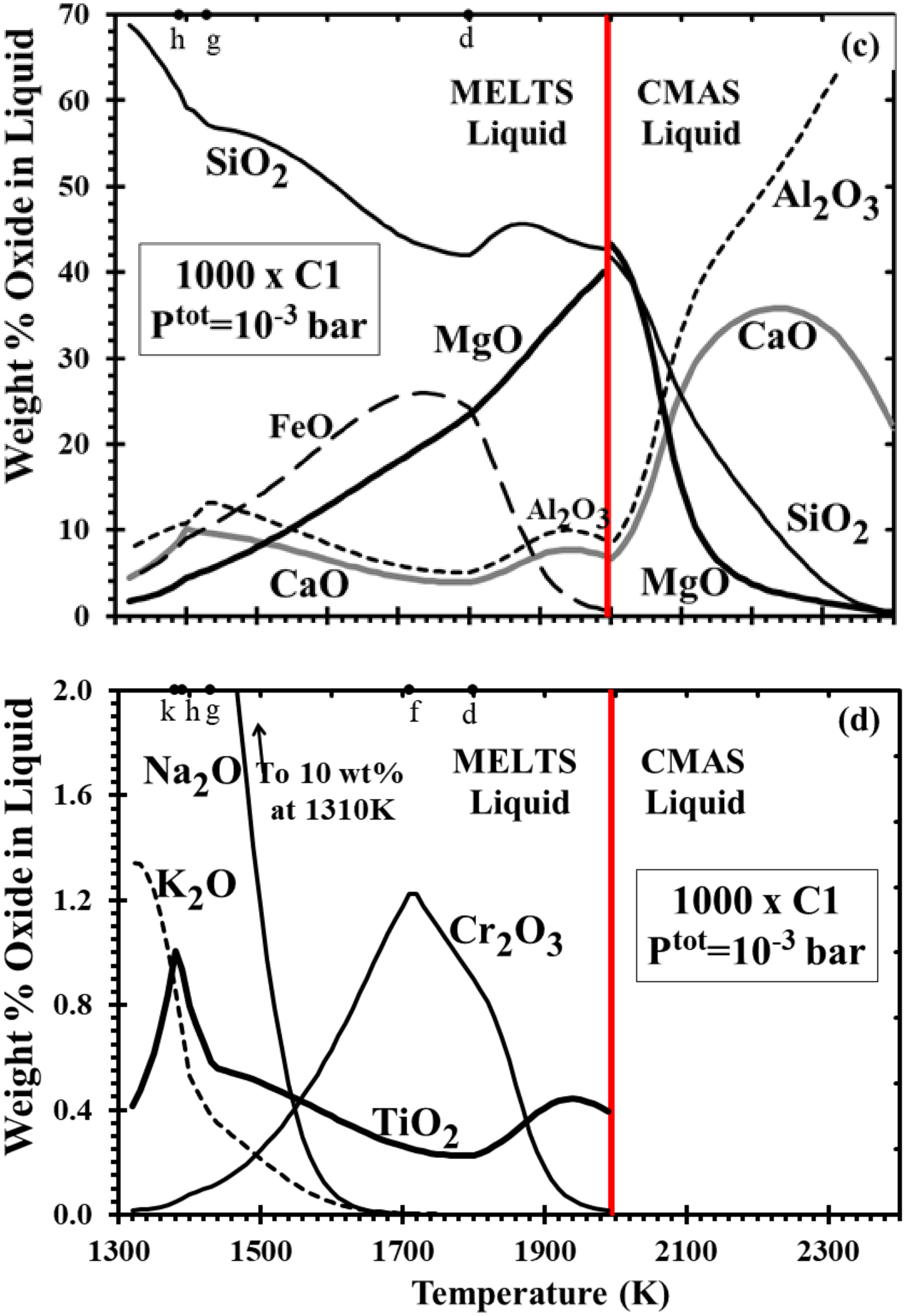

**Figure 10:** (continued)

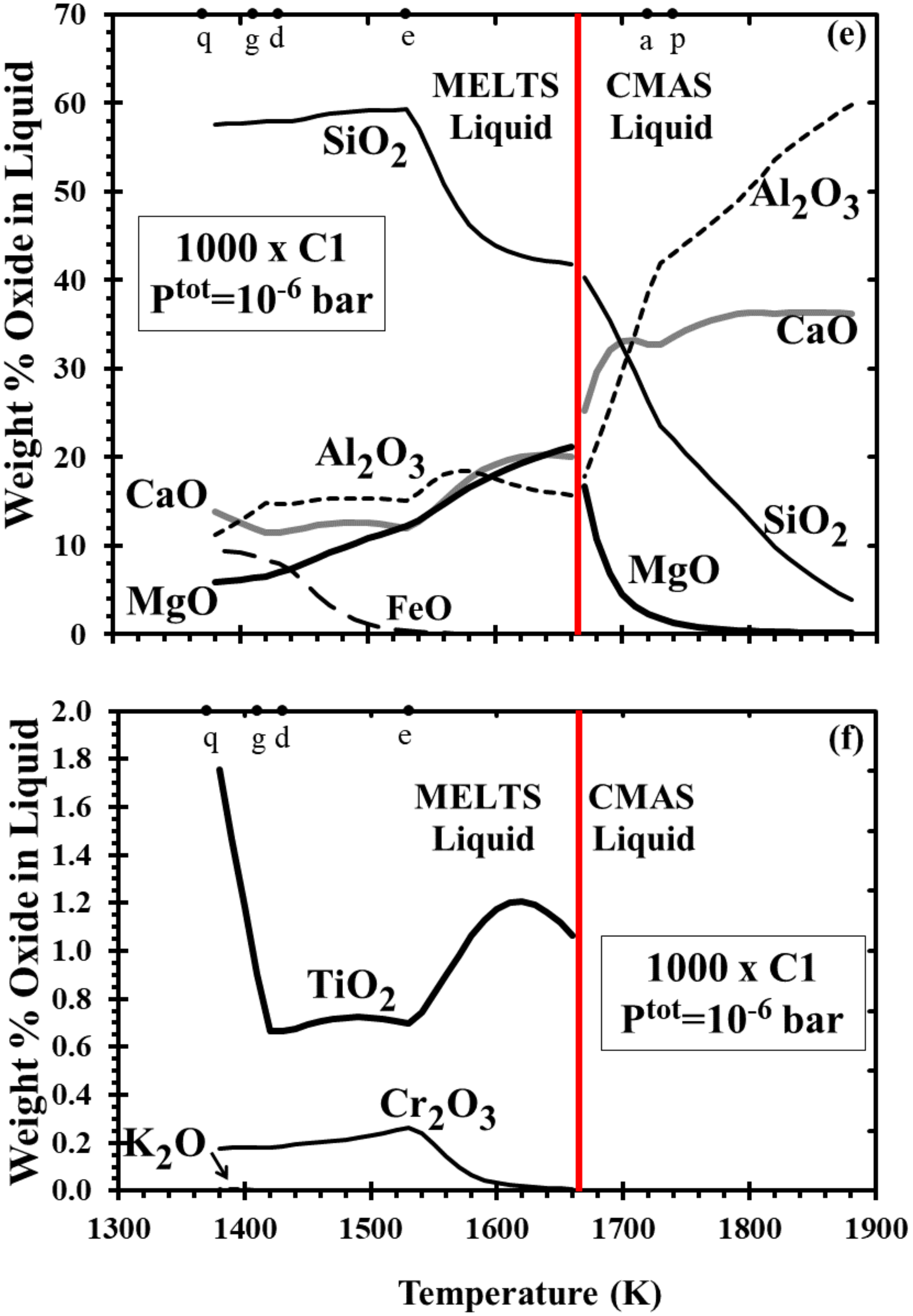

**Figure 10:** (continued)

Below the condensation temperature of olivine, the main difference in major element trends of the liquid at different combinations of $P^{tot}$ and dust enrichment, aside from those of FeO and alkalis, is the failure of the $SiO_2$ concentration to level off with falling temperature at a dust enrichment of 1000x at $10^{-3}$ bar (Fig. 10c) as it does at a dust enrichment of 100x at $10^{-3}$ bar (Fig. 10a) and 1000x at $10^{-6}$ bar (Fig. 10e). This is entirely due to the absence of orthopyroxene from the crystalline assemblage in equilibrium with the liquid at a dust enrichment of 1000x at $10^{-3}$ bar. It is the condensation of this phase which triggers the flattening of the $SiO_2$ curve under the other sets of conditions. Of the three cases shown, it is at a dust enrichment of 1000x and $10^{-3}$ bar where the $Cr_2O_3$ content of the liquid is highest, climbing to 1.2 wt% with falling temperature and then declining after Cr-spinel becomes stable at 1710K (Fig. 10d). At a dust enrichment of 100x at $10^{-3}$ bar, this phase becomes stable at 1600K, before the $Cr_2O_3$ content of the liquid reaches 0.36 wt% (Fig. 10b). At a dust enrichment of 1000x at $10^{-6}$ bar, Cr-spinel coexists with the MELTS liquid over its entire stability range, preventing its $Cr_2O_3$ content from exceeding 0.28 wt% (Fig. 10f). As the amount of liquid becomes vanishingly small during near-solidus crystallization of clinopyroxene and plagioclase, concentrations of $TiO_2$ are seen to build up in the last dregs of liquid. Only at a dust enrichment of 1000x at $10^{-3}$ bar does this trend reverse itself. This is due to stabilization of a pyrophanite-rich solid solution at a temperature above that for the disappearance of liquid. As seen by comparing Figs. 10b, d and f, alkali contents of the liquid increase both with increasing $P^{tot}$ and with increasing dust enrichment because the partial pressures of sodium and potassium increase with both parameters. As a result, $Na_2O$ and $K_2O$ concentrations in the liquid are negligible at $10^{-6}$ bar, even at a dust enrichment of 1000x. In the other cases shown, $Na_2O$ and $K_2O$ concentrations rise above negligible levels only within 100-200K of the temperature of disappearance of liquid, reaching maxima of 10.1 wt% and 1.3 wt%, respectively, at $10^{-3}$ bar and a dust enrichment of 1000x. At a dust enrichment of 1000x, the FeO content of the liquid at $10^{-3}$ bar is higher than at $10^{-6}$ bar at most temperatures (Figs. 10c and 10e), considerably so at some temperatures. Because $f_{O_2}$ is only weakly dependent on $P^{tot}$ at 1500-1600K, the higher FeO content of the liquid is due simply to the higher $P_{Fe}$ at higher $P^{tot}$, which causes a greater proportion of the iron to be condensed at any given temperature.

In most cases, the liquid disappears in the temperature interval 1370 to 1400K, the approximate location of the peridotite solidus at 1 bar (see Table 5). An exception to this general rule is found in Table 8 for the case of a dust enrichment of 1000x at $10^{-3}$ bar, where the liquid persists to 1310K. At the same $P^{tot}$ and a dust enrichment of 500x, the liquid disappears at a significantly higher temperature, 1400K. Similarly, at the same dust enrichment, 1000x, and lower $P^{tot}$, $10^{-6}$ bar, the liquid also disappears at a much higher temperature, 1370K. At $10^{-3}$ bar, the reason for the different solidification temperatures at the different dust enrichments is evident from a comparison of the liquid compositions in the two cases at 1410K, the last temperature step before the liquid disappears at a dust enrichment of 500x. At this temperature, the liquid at the lower dust enrichment contains slightly less $Na_2O$, 2.67 wt%, and much less FeO, 3.26%, than the liquid at the higher dust enrichment, 3.81 and 9.46%, respectively, and high concentrations of both of these oxides are known to depress solidus temperatures. At a dust enrichment of 1000x, the reason for the different solidification temperatures at the different total pressures is found in the different liquid compositions at 1380K, the last temperature step before the liquid disappears at $10^{-6}$ bar. Although the FeO content of the liquid is slightly lower at $10^{-3}$ than at $10^{-6}$ bar, 7.54 *vs* 9.51 wt%, the $Na_2O$ concentration is much higher at $10^{-3}$ than at $10^{-6}$ bar, 5.89 *vs* <0.01 wt%. This is because the partial pressure of Na is more than a factor of 200 higher at 1380K at $10^{-3}$ bar than at $10^{-6}$ bar. Furthermore, because Na continues to condense into the liquid in this temperature range, the lower the temperature to which the liquid persists, the higher its $Na_2O$ content becomes, and this further lowers the ultimate temperature of its disappearance.

### 3.8. Composition of Spinel

The numbers of cations in spinel per 4 oxygen atoms are plotted as a function of temperature at dust enrichments of 100x and 1000x at $10^{-3}$ bar in Figs. 11a and 11b, respectively, and 1000x at $10^{-6}$ bar in Fig. 11c. In all cases, the highest temperature spinel forms by reaction of gaseous Mg with $TiO_2$ in perovskite and $Al_2O_3$ in the CMAS liquid, except at 1000x and $10^{-3}$ bar, where all Ti is from the gas. In this spinel, the Ti cations first increase with falling temperature as perovskite and/or gaseous Ti are consumed and then decrease sharply when the MELTS liquid, which can accommodate Ti, becomes stable. Both stages proceed in accordance with the coupled substitution of $Mg^{2+} + Ti^{4+} = 2Al^{3+}$ and are accompanied by steadily rising numbers of Fe and Cr cations which are condensing from the gas. As discussed previously, the high Ti contents of these spinels and possibly even their existence, may be artifacts of the inability of the CMAS liquid to accommodate Ti. The incoming of the MELTS liquid causes the very Ti-rich spinel at 1000x and $10^{-3}$ bar to dissolve suddenly, and the less Ti-rich spinel at 100x and $10^{-3}$ bar to dissolve gradually before disappearing. The even lower-Ti spinel at 1000x and $10^{-6}$ bar continues to crystallize with falling temperature, gradually becoming first more Cr-rich and then more Fe-rich. As shown in Fig. 11c, the Cr/Al ratio levels off below 1390K, as formation of spinel continues by reaction of gaseous Cr with $Al_2O_3$ in the liquid. When spinel re-forms in the two cases at $10^{-3}$ bar, its Cr/Al ratio falls, as the spinel draws down $Al_2O_3$ from the liquid while deriving its Cr from the metal alloy and the gas at 100x, and from the metal alloy and the liquid at 1000x. In all three cases, these trends are interrupted by plagioclase formation, which draws its $Al_2O_3$ from the $MgAl_2O_4$ component of the spinel, increasing the Cr/Al ratio and decreasing the amount of spinel. Plagioclase formation also causes an increase in the rate of increase of the number of Ti cations in the spinel with decreasing temperature, accompanied by an increase in the number of Mg and/or Fe ions in accordance with the above coupled substitution. At lower temperature, the number of Ti ions in the spinel begins to decrease with decreasing temperature due to extraction of Ti into pyrophanite or, at 100x and $10^{-3}$ bar, clinopyroxene.

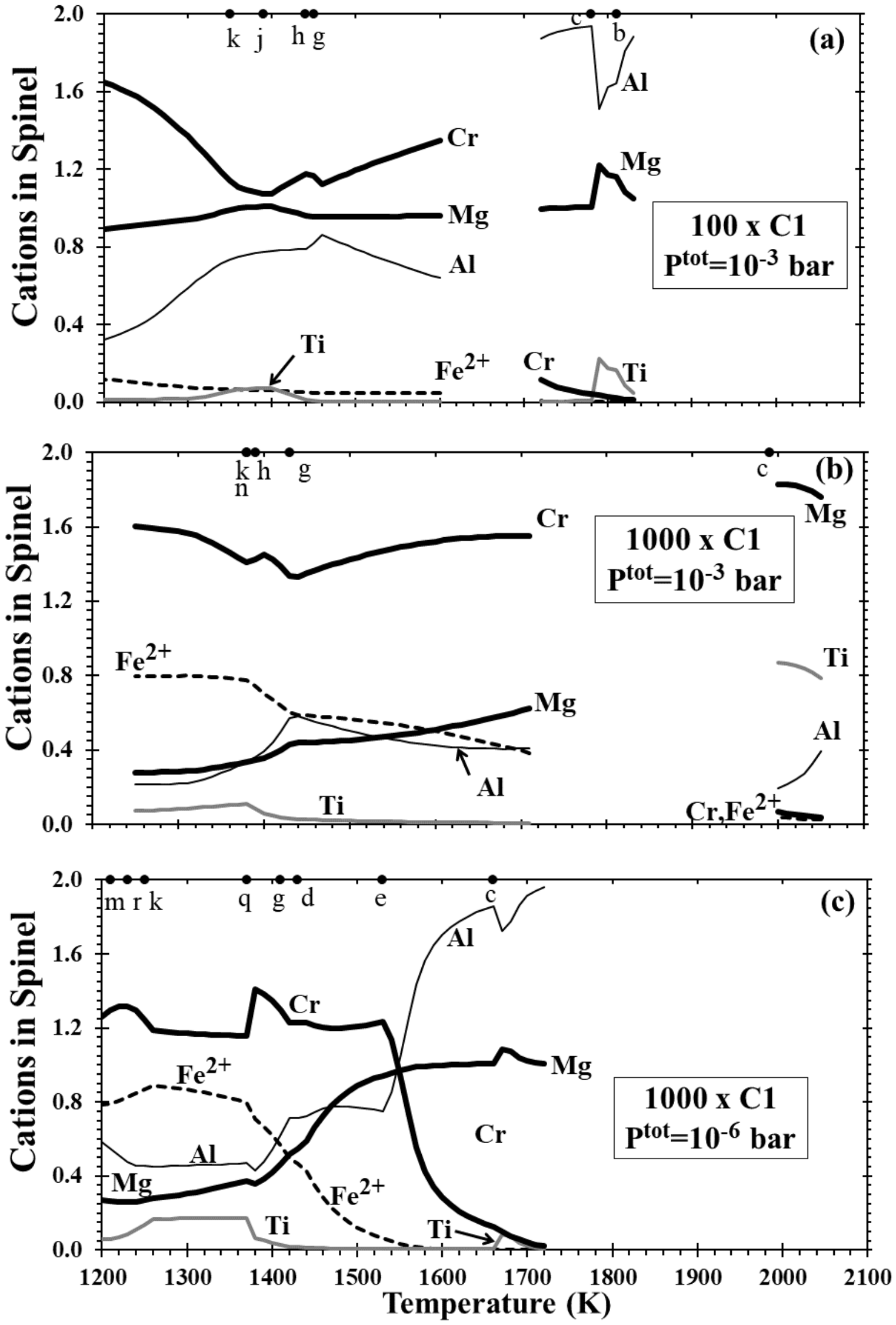

**Figure 11:** Composition of spinel as a function of temperature at (a) $P^{tot} = 10^{-3}$ bar and a dust enrichment of 100x; (b) $P^{tot} = 10^{-3}$ bar and a dust enrichment of 1000x; and (c) $P^{tot} = 10^{-6}$ bar and a dust enrichment of 1000x. Inflection points labelled as in Fig. 10, plus: b, perovskite out; c, olivine in; j, liquid out; m, MnO in; n, pyrrhotite in; r, orthopyroxene out.

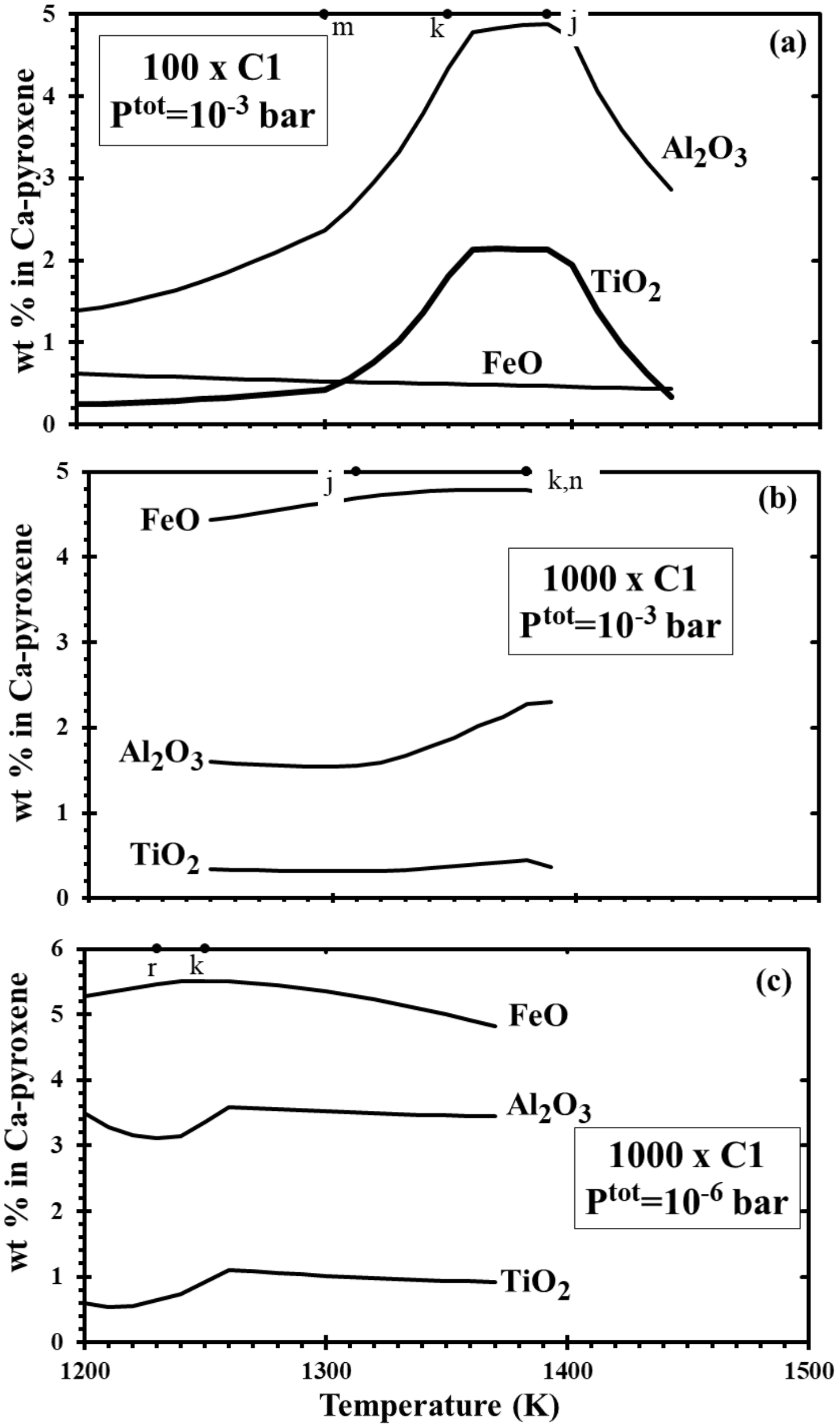

**Figure 12:** Composition of Ca-rich clinopyroxene as a function of temperature at (a) $P^{tot} = 10^{-3}$ bar and a dust enrichment of 100x; (b) $P^{tot} = 10^{-3}$ bar and a dust enrichment of 1000x; and (c) $P^{tot} = 10^{-6}$ bar and a dust enrichment of 1000x. Inflection points as previously labelled.

### 3.9. Composition of Clinopyroxene

The concentrations of FeO, $Al_2O_3$ and $TiO_2$ in clinopyroxene are plotted as a function of temperature at $10^{-3}$ bar and dust enrichments of 100x and 1000x, and at $10^{-6}$ bar and 1000x in Figs. 12a, b and c, respectively. The amount of clinopyroxene increases with falling temperature in all three cases due either to crystallization from the liquid or, after liquid is exhausted, to reactions among plagioclase, orthopyroxene and olivine, as can be seen in Figs. 6 and 7 for the cases at $10^{-3}$ bar. The proportion of the total Fe accounted for by clinopyroxene increases with falling temperature as metal is oxidized, but the concentration of FeO may rise or fall depending on the relative rates of formation of Mg and Fe endmembers. Similarly, the proportions of the total Al and Ti accounted for by clinopyroxene increase with falling temperature as this phase crystallizes from the liquid in the cases at $10^{-3}$ bar, but the concentrations of $Al_2O_3$ and $TiO_2$ may increase or decrease with falling temperature due to the relative formation rates of the different pyroxene endmembers. At $10^{-3}$ bar and 100x and at $10^{-6}$ bar and 1000x, a temperature is reached below which the $Al_2O_3$ and $TiO_2$ concentrations begin to fall with decreasing temperature, as plagioclase begins to draw its $Al_2O_3$, and pyrophanite its $TiO_2$, from clinopyroxene.

### 3.10. Composition of Feldspar

The mole fractions of albite and orthoclase in feldspar are plotted as a function of temperature at $10^{-3}$ bar and dust enrichments of 100x and 1000x in Fig. 13. The amount of feldspar and its albite and orthoclase contents increase steadily with decreasing temperature in both cases. Above the temperature of disappearance of liquid, feldspar draws its Na from both liquid and gas but its K from the liquid only. Below this temperature, Na and K continue to condense from the gas into feldspar, increasing their concentrations in feldspar with decreasing temperature. At $10^{-6}$ bar and dust enrichments of 100x and 1000x, K contents of feldspar are vanishingly small down to the last temperature step of the computations shown on Table 7. This is also true for Na at 100x, but $X_{Ab}$ at 1000x rises to 0.1 at 1200K.

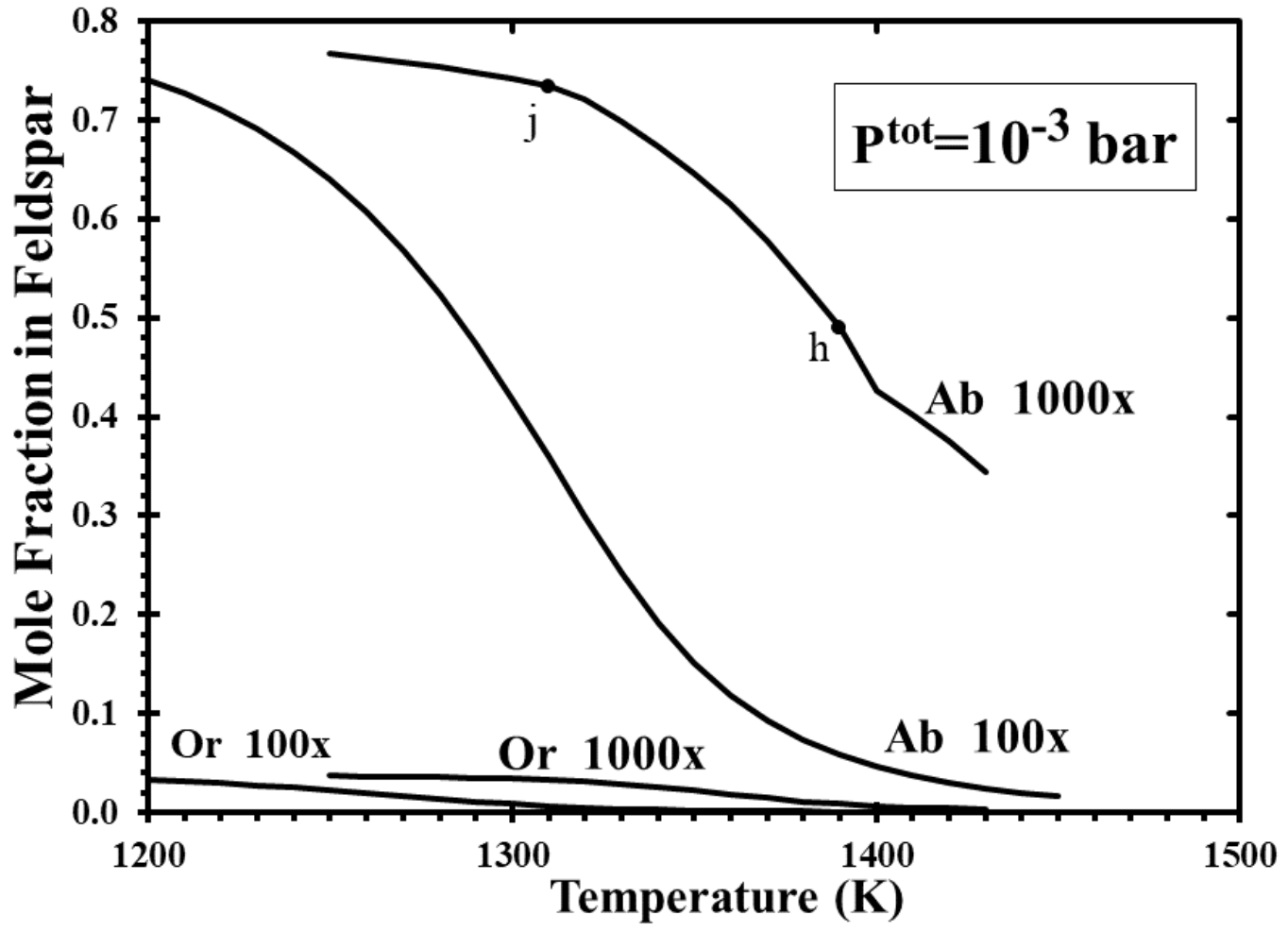

**Figure 13:** Mole fractions of albite (Ab) and orthoclase (Or) in feldspar as a function of temperature at $P^{tot} = 10^{-3}$ bar and a dust enrichment of 100x and 1000x. Inflection points as previously labelled.

### 3.11. Composition of Metallic Nickel-Iron

The concentrations of Ni, Co and Cr in the metallic nickel-iron alloy at various combinations of $P^{tot}$ and dust enrichment are shown in Figs. 14a, b and c, respectively. Under all conditions shown, Ni and Co are slightly more refractory and have slightly steeper condensation curves than Fe. This leads to high concentrations of Ni and Co in the first-condensing alloys, steadily declining concentrations of Ni and Co with falling temperature as condensation of slightly less refractory Fe dilutes the previously condensed Ni and Co, and finally a leveling off of the Ni and Co contents when all three elements are totally condensed. At still lower temperatures in the more oxidizing cases, 1000x at $10^{-3}$ and $10^{-6}$ bar, Ni and Co contents begin to rise very gradually with falling temperature due to oxidation of the Fe component of the alloy. At 1380K at 1000x and $10^{-3}$ bar, the Ni and Co contents of the alloy begin to rise very sharply with falling temperature due to reaction of gaseous sulfur with the Fe component of the alloy to form pyrrhotite. Under oxidizing conditions, Cr is slightly more refractory than Fe and, like Ni and Co, falls steadily in concentration with falling temperature. Under more reducing conditions, however, the behavior of Cr is completely different. At 100x and $10^{-3}$ bar, Cr is slightly less refractory than Fe, its concentration in the metal increases sharply with falling temperature in the high-temperature alloys, and only reverses itself below the formation temperature of Cr-spinel, which extracts Cr from the metal alloy. The increase in Cr content with falling temperature is not seen at 100x and $10^{-6}$ bar because most of the Cr has already condensed as Cr-spinel at a higher temperature than that where the metal alloy begins to condense. While high Si concentrations in metallic nickel-iron alloys can result from condensation from gases more reducing than a gas of solar composition, $X_{Si}$ is always < $10^{-4}$ in the systems considered in this work.

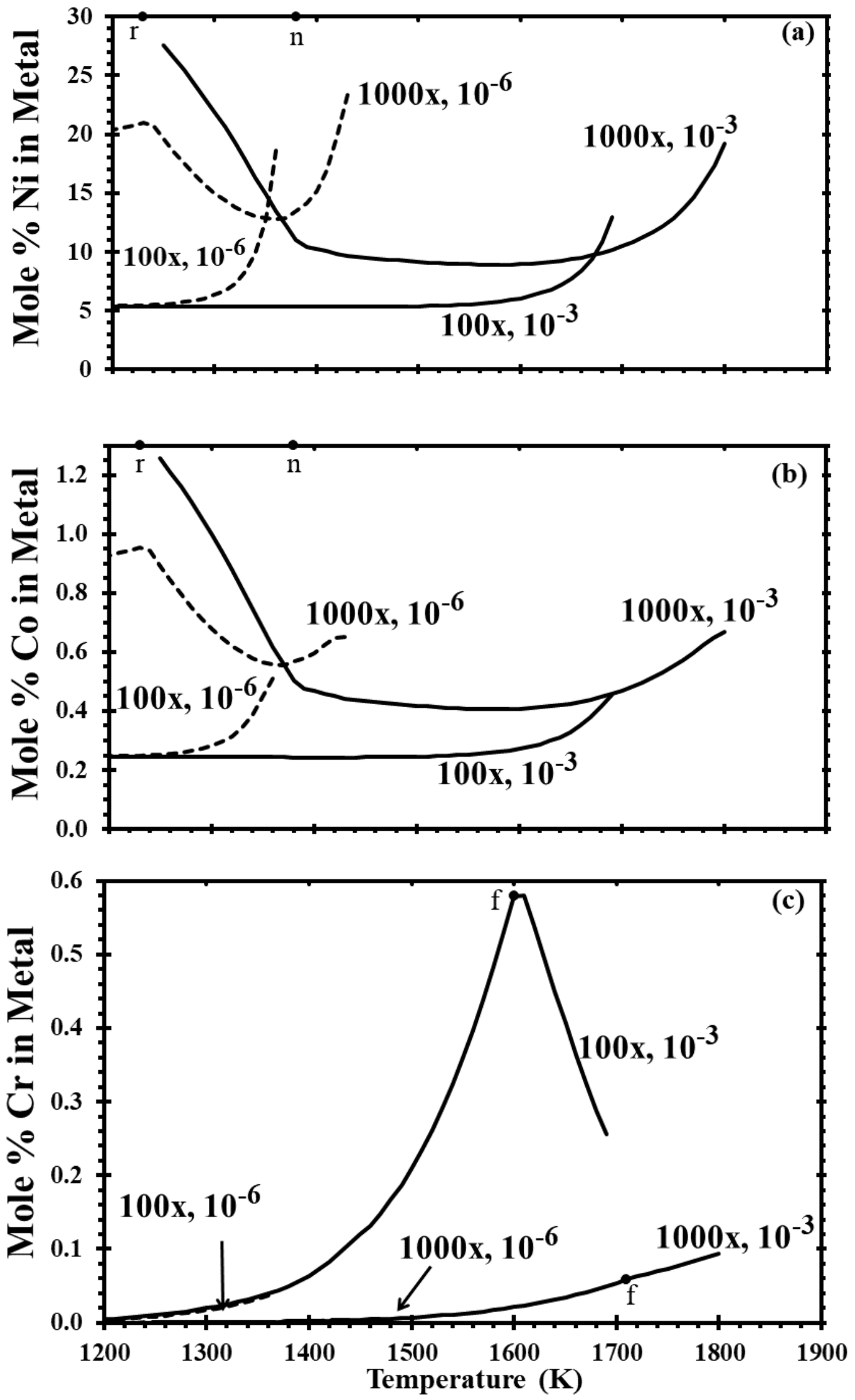

**Figure 14:** Mole per cent of (a) Ni; (b) Co; and (c) Cr in metallic nickel-iron alloy as a function of temperature at the stated conditions. Inflection points as previously labelled.

### 3.12. Metal-Sulfide Condensate Assemblages

Because of the high concentration of sulfur in dust of C1 composition, enrichment in such dust leads to much higher $f_{S_2}$ and permits sulfide phases to condense at higher temperatures at a given $P^{tot}$ than in a gas of solar composition. Inspection of Table 7 reveals that no sulfide phase becomes stable above the last temperature step of the calculations at any of the dust enrichments shown at $10^{-6}$ bar, but that, at $10^{-3}$ bar, pyrrhotite, $Fe_{0.877}S$, joins metallic nickel-iron as a stable condensate at 1330 and 1380K at dust enrichments of 500x and 1000x, respectively. These temperatures are higher than minimum melting temperatures in the Ni-poor part of the Fe-Ni-S system but, since our computer program does not contain a thermodynamic model for Fe-Ni-S liquids, it would be unable to predict their existence even if they were more stable than the metal + pyrrhotite assemblages that are predicted. In order to see if sulfide liquids are more stable than the predicted assemblages, the relative atom proportions of Fe and S were calculated for the metallic nickel-iron + pyrrhotite assemblage predicted at each temperature step for dust enrichments of 500x, 800x and 1000x at $10^{-3}$ bar, and are plotted on a portion of the liquid-crystal phase relations in the Fe-S binary (Chuang *et al*., 1986a) in Fig. 15. The dashed curves in this figure are the phase boundaries that result from addition of 7% Ni to the system, taken from the work of Hsieh *et al*. (1982), projected onto the Fe-S plane.

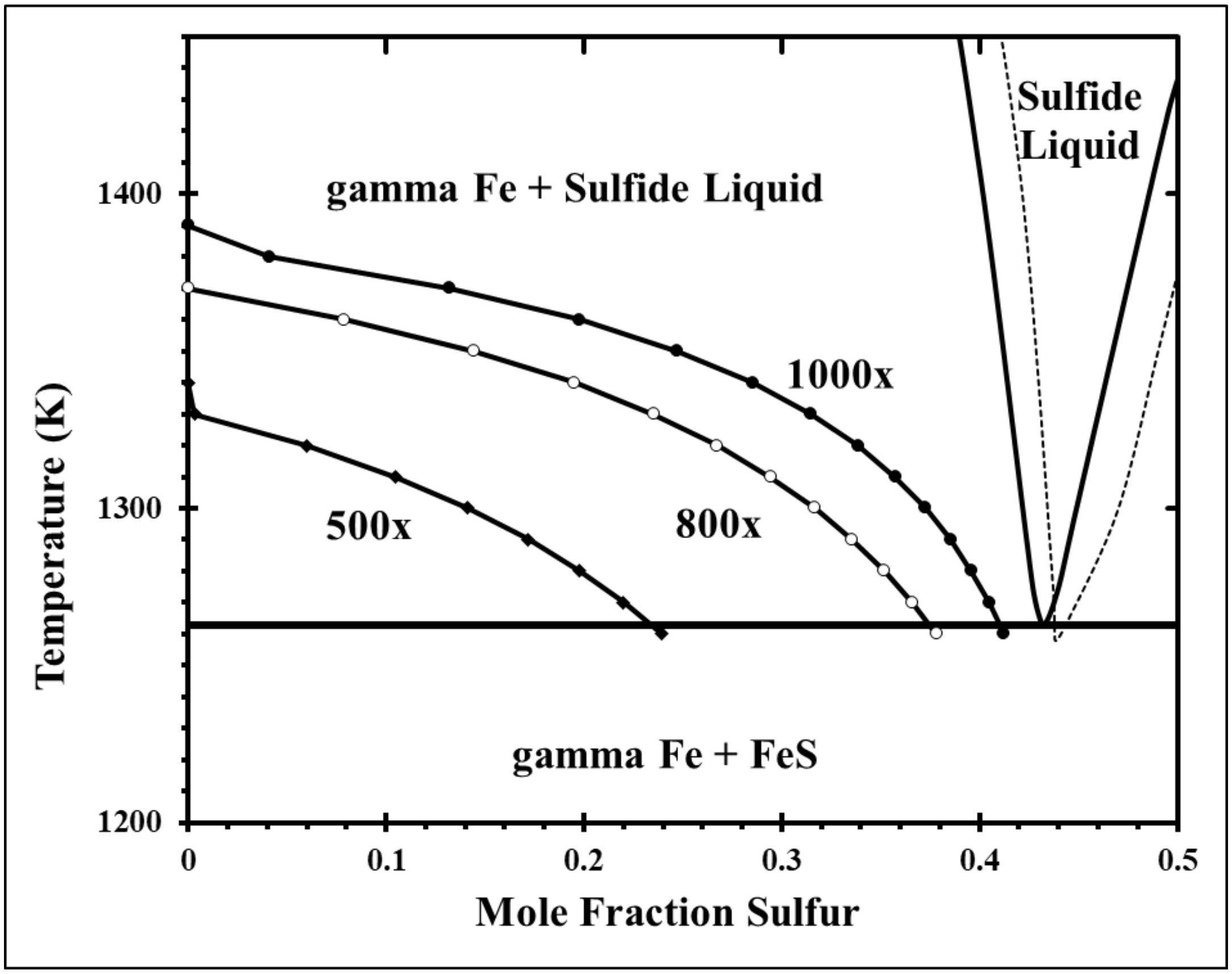

**Figure 15:** Bulk chemical compositions of metallic nickel-iron + pyrrhotite condensate assemblages predicted at $P^{tot}=10^{-3}$ bar and the stated dust enrichments, projected onto the liquid-crystal phase relations of the Fe-rich portion of the Fe-S binary system. Dashed curves show projections of phase boundaries when 7 mole % Ni is present.

Under all conditions, the trajectory of the condensate compositions initially falls vertically along the left margin of the diagram until the temperature of pyrrhotite formation is reached. Below this, the trajectories extend to the right and downward, well above the eutectic temperature of 1262K and well within the field of gamma-iron + Fe-S liquid. The solid assemblage of metallic nickel-iron + pyrrhotite predicted by our thermodynamic model to condense under these conditions is thus seen to form at temperatures where it is actually metastable relative to metallic iron + Fe-S liquid. Nickel, cobalt and chromium are also predicted to condense into the metal alloy, but the amounts of Co and Cr are quite small, and the phase relations are seen to change very little with the addition of 7% Ni, a fairly representative concentration in these condensate assemblages. We conclude that, at these relatively high total pressures and dust enrichments, direct condensation of iron sulfide liquids will occur. Furthermore, because the liquid-bearing assemblage obviously has a lower Gibbs free energy than the predicted assemblage, the temperature of appearance of sulfide liquid will actually be higher than the condensation temperature of pyrrhotite. Recall that the silicate liquid in the 500x case solidifies at 1400K, which is well above the minimum condensation temperature of sulfide liquid, but that the silicate liquid at 1000x does not solidify until 1310K, almost 100K below the minimum condensation temperature of sulfide liquid in this case. This means that, at the highest dust enrichment factors at high $P^{tot}$, condensation of coexisting silicate and sulfide liquids occurs, assuming they are not miscible.

## 4. DISCUSSION

### 4.1. Stability of Silicate Liquid in Solar Gas

The calculations show that no liquids are stable in a gas of solar composition, even at a total pressure as high as $10^{-3}$ bar. Any liquids formed by the partial or complete melting of agglomerated solids (Whipple, 1966; Lofgren, 1996) would therefore be highly unstable with respect to partial evaporation in a solar nebula of canonical composition, and would become even more unstable with decreasing pressure. Such liquids would lose FeO and alkali metals most readily, followed by Mg and Si, then Ca and Al with increasing temperature.

Experimentally determined evaporation rates of $Na_2O$ from liquids of chondrule composition (Radomsky and Hewins, 1990; Tsuchiyama *et al.*, 1981; Yu and Hewins, 1998) show that Na loss is faster at lower total pressures and lower $f_{O_2}$. However, Lewis *et al.* (1993) showed that sodium loss in alkali olivine basalt melt droplets (3.05 wt% $Na_2O$) was nearly attenuated upon heating above 1600K at $P^{tot}$=1 bar, at the iron-wüstite buffer in a $CO/CO_2$ + NaCl vapor with $P_{Na} > 4x10^{-6}$ atm. The present calculations show that these conditions are roughly equivalent to a C1 dust enrichment factor well in excess of 1000x at $10^{-3}$ bar. Lewis *et al*. (1993), based on their experiments and following Wood (1984), suggested chondrule formation in 'clumps' where the local partial pressures of condensable elements and oxygen were enhanced by volatilization of chondrule precursor material. The phase diagrams presented above provide the rigorous thermochemical basis for concluding that such a mechanism would indeed stabilize liquids in the solar nebula and reduce or eliminate the driving force for volatilization of Na and other elements from chondrule-like liquids.

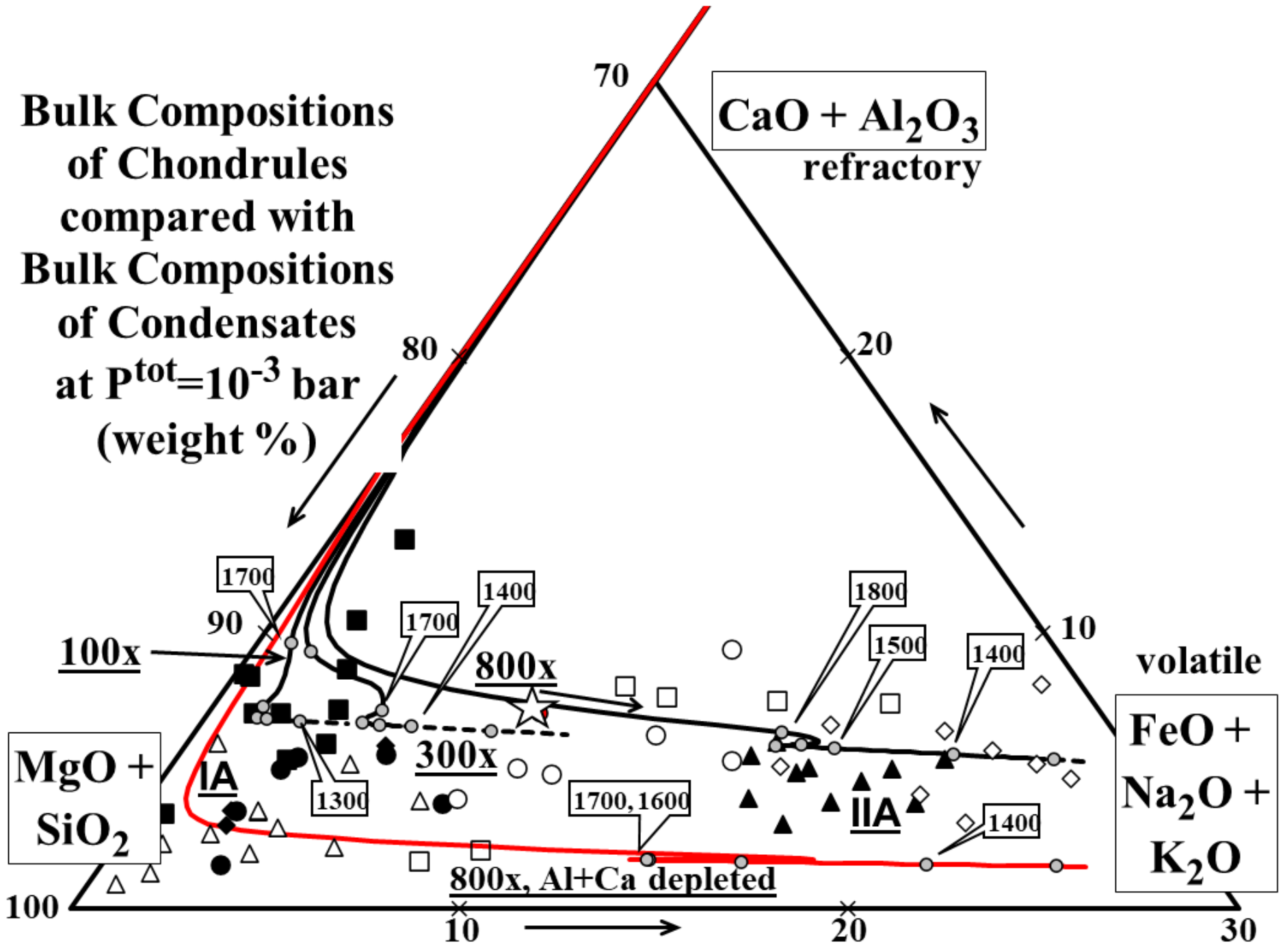

**Figure 16:** Bulk compositions of chondrules, and bulk compositions of condensed oxides at $10^{-3}$ bar. Data: filled squares=Type IA (Jones and Scott, 1989); filled diamonds=Type IB (Jones, 1994); filled circles=Type IAB (Jones, 1994); filled triangles=Type IIA (Jones, 1990); open squares=Type IIB (Jones, 1996); open circles=H3 (Lux *et al*., 1981); open triangles=CM and CO (Rubin & Wasson, 1986); open diamonds=Manych L3 (Dodd, 1978). Circles on each path correspond to compositions at 1800, 1700, 1600, 1500, 1400, and 1300K with increasing FeO + $Na_2O$. Dashed extensions of paths are subsolidus compositions to 1200K for 100x and 300x and to 1250K for 800x. Star represents the bulk composition of peridotite KLB-1 (Takahashi, 1986).

### 4.2. Chondrules in Dust-enriched Systems

Is there some pressure and dust enrichment at which the temperature variation of the bulk composition of the condensed matter, exclusive of metal, resembles the composition range of chondrules? For example, are chondrules the quenched droplets of the solid + liquid assemblages that formed by equilibrium condensation at various temperatures? In Fig. 16, chondrule compositions spanning the common composition range are plotted in the forsterite-rich corner of the ternary ($CaO + Al_2O_3$) -- ($MgO + SiO_2$ ) -- ($FeO + Na_2O + K_2O$). This perspective is useful because the apices correspond to groups of oxides which condense in distinct temperature ranges, and these seven oxides constitute the bulk of chondritic material. Superposed on these chondrule compositions are paths representative of the trajectories of bulk oxide condensates, exclusive of metal, with decreasing temperature. Each of the condensation paths sweeps down from the $CaO + Al_2O_3$ apex toward the field occupied by Type IA chondrules (Jones and Scott, 1989), close to the $MgO + SiO_2$ corner. Note that no single path for a particular combination of $P^{tot}$ and dust enrichment will be able to account for all of the chondrule compositions plotted. Of the cases shown, only the one at the highest dust enrichment, 800x, is oxidizing enough to make sufficient FeO at high temperature that the condensation trajectory passes through the field of Type IIA (Jones, 1990) chondrules at temperatures above that where dust and gas could be expected to equilibrate, *i.e.* ~1200K. This case is so oxidizing, however, that significant FeO condenses before complete condensation of MgO and $SiO_2$, causing the trajectory to peel away from the $CaO + Al_2O_3$—$MgO + SiO_2$ join before reaching the $MgO + SiO_2$ corner and thus only to graze the top of the field of Type IA chondrules. The condensation paths at lower dust enrichments, 100x and 300x, are seen to penetrate the Type IA field only slightly more, missing the vast majority of the data points in it as well as the Type IIA field. Similar composition trajectories to these are obtained at $P^{tot} < 10^{-3}$ bar but, at a given dust enrichment, the temperatures at which the trajectories reach high FeO + alkali contents become progressively lower with decreasing $P^{tot}$. Sequestration of high-temperature, Ca-, Al-rich condensates could cause the condensation path of the remaining system at a particular $P^{tot}$ and dust enrichment to pass through the middle of the field of Type IA chondrules, but such a path would then graze only the bottom of the field of Type IIA chondrules. An example of such a path is shown in Fig. 16 for the case of condensation of a system having a dust enrichment of 800x but from which 72 % of the $Al_2O_3$ and CaO have been removed. Given the wide range of $CaO + Al_2O_3$ contents of the chondrules in Fig. 16, it is conceivable that a family of condensation curves corresponding to a single $P^{tot}$ and a high dust enrichment but variable amounts of sequestration of high-temperature condensates could account for the observed data. Although such an appeal to a multitude of adjustable parameters appears to satisfy the data, the separate roles of FeO and alkalies are not addressed in Fig. 16.

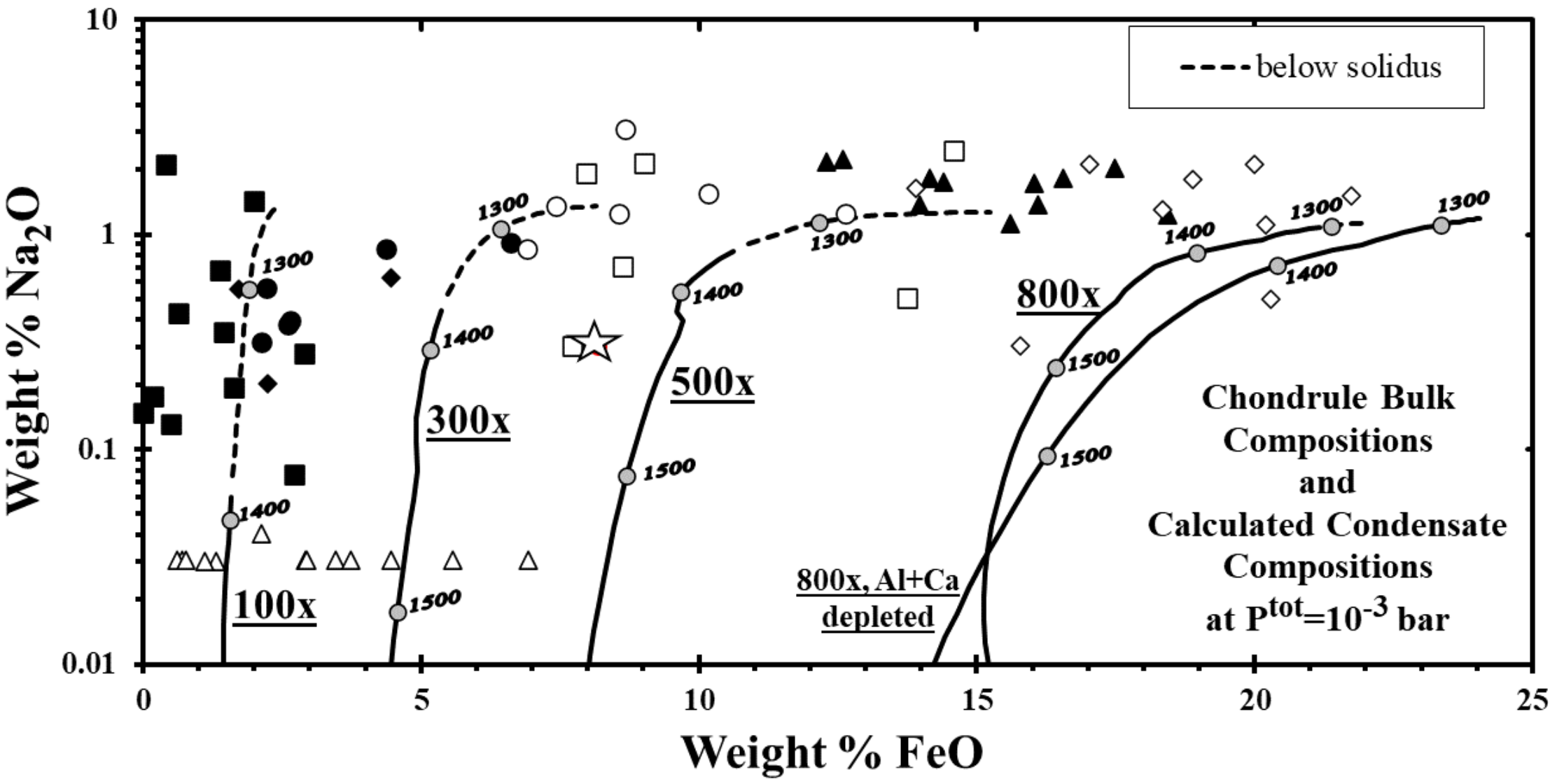

**Figure 17:** Sodium and iron oxide contents of the chondrules of Fig. 16 and of KLB-1, with representative paths of bulk oxide condensates. Compositions reported as ≤ 0.03% $Na_2O$ are plotted at 0.03%.

The FeO and $Na_2O$ contents of the same chondrules shown in Fig. 16 are illustrated in Fig. 17. Representative oxide bulk compositions for condensation paths at 100x, 300x, 500x, and 800x dust enrichment, the latter with and without removal of 72 % of the $Al_2O_3$ and CaO, at $10^{-3}$ bar are superposed on Fig. 17, with the subsolidus portions dashed. The FeO contents of the calculated condensate assemblages at high dust enrichments reach the levels found in chondrules at high temperatures, but $Na_2O$ contents only approach the levels found in Na-rich chondrules near the solidus of silicate liquid. The compositions of the chondrules richest in both $Na_2O$ and FeO, which include the Type IIA and IIB chondrules of Jones (1990; 1996) and many of those reported by Dodd (1978), are more $Na_2O$-rich than compositions on any of the equilibrium condensation paths calculated under the conditions investigated here. This result precludes formation of these particular chondrules by representative sampling of equilibrium condensate assemblages from dust-enriched systems at specific temperatures, either by quenching and isolating primary condensates or by isochemical melting and quenching of such samples at $P^{tot} \leq 10^{-3}$ bar and dust enrichments ≤1000x. Dust enrichment factors of >800x are not required, however, to produce the iron contents observed in most chondrules. If the FeO contents of chondrules *did* result from the formation and isolation of their precursors in a dust-enriched environment, their Na contents may have resulted from some secondary, as yet unspecified process.

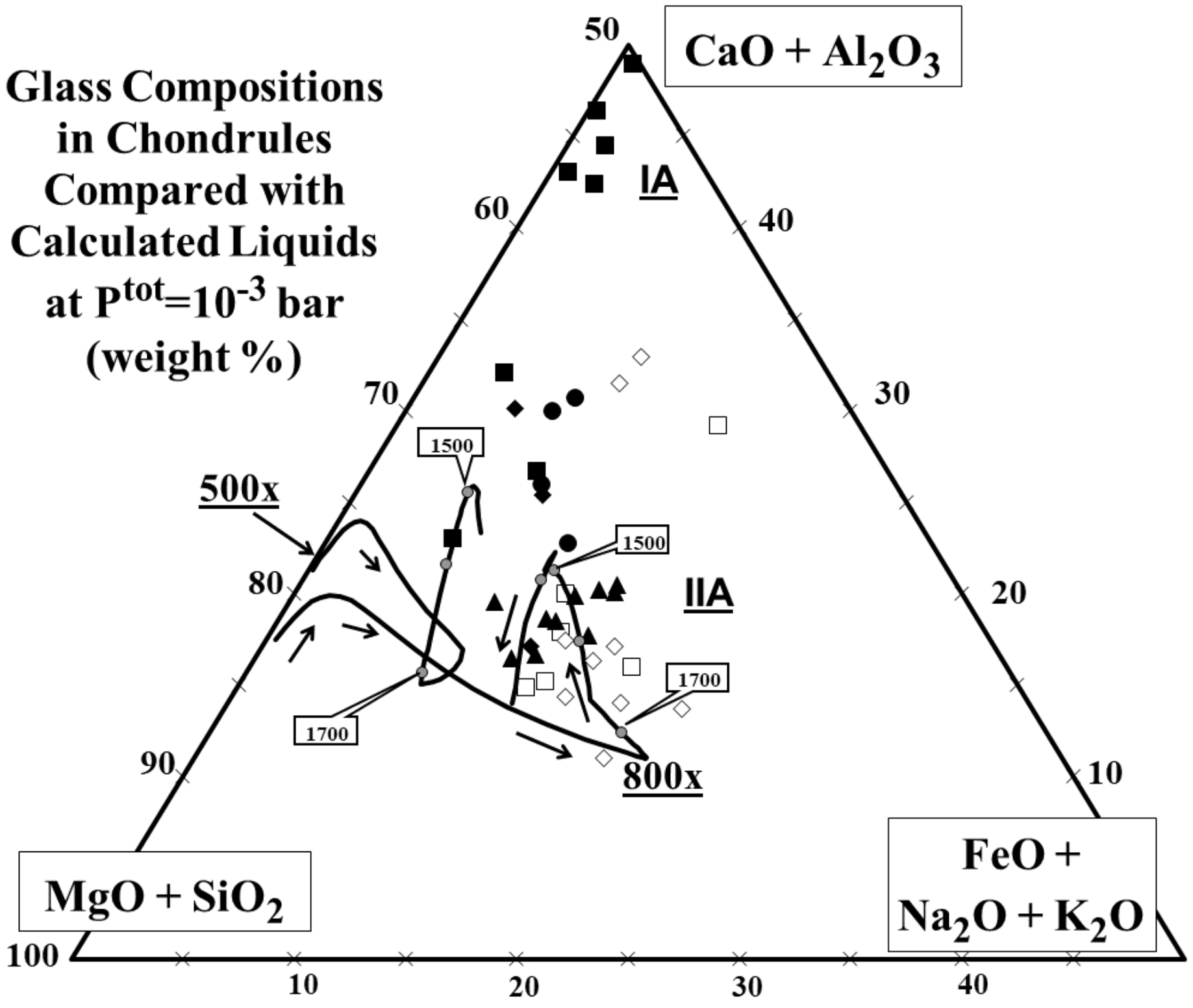

**Figure 18:** Glass compositions in chondrules, with paths of condensate liquids for 500 and 800x C1 dust enrichment, from below the appearance temperatures of olivine to their solidi. Symbols for chondrule data as in Fig. 16. Grey circles on each path mark hundred-degree decrements starting at 1700K.

Evidence for only limited intra-crystal and liquid-crystal equilibrium in chondrules includes the presence of grains interpreted to be relict olivine and/or pyroxene (Steele, 1986) and chemical zoning in phenocrysts of porphyritic chondrules (Simon and Haggerty, 1979; Jones and Scott, 1989). Because diffusion of major elements in silicate liquids is generally much faster than through crystals, the liquid is likely to have been the last phase to have even partially equilibrated with ambient vapor, and the glass is thus the most likely phase to have preserved a record of the temperature, pressure and composition of the vapor. In fact, radial chemical zonation in mesostasis of unequilibrated chondrules, particularly with respect to $Na_2O$ (Ikeda and Kimura, 1985; DeHart *et al.*, 1988; Grossman *et al.*, 1997) suggests that some chondrules underwent only partial equilibration at a late stage in their formation, possibly with surrounding gas. Are there conditions for which the liquids in equilibrium with dust-enriched vapor have the same compositions as the glass found in chondrules? Plotted on Fig. 18 are the compositions of the mesostasis from most of the chondrules whose bulk compositions are illustrated in the previous two figures. The axes of Fig. 18 are the same as those of Fig. 16, and the composition trajectories for the liquid fraction of the condensates at two dust enrichments are superposed on Fig. 18, starting at the temperature where olivine first condenses, and the transition from CMAS to MELTS liquid models occurs. The calculated liquids move away from forsterite toward the CaO + $Al_2O_3$ apex, then begin to become enriched in iron. Once past the peak in iron content, the paths differ in trajectory, due to differences in the proportions of crystallizing olivine and orthopyroxene, and liquid. Continuing olivine and pyroxene crystallization drives the liquid trajectory steeply away from the MgO + $SiO_2$ corner, causing CaO and $Al_2O_3$ concentrations to rise in the liquid until feldspar and pyroxene crystallize, very near the solidus, driving CaO and $Al_2O_3$ downward again.

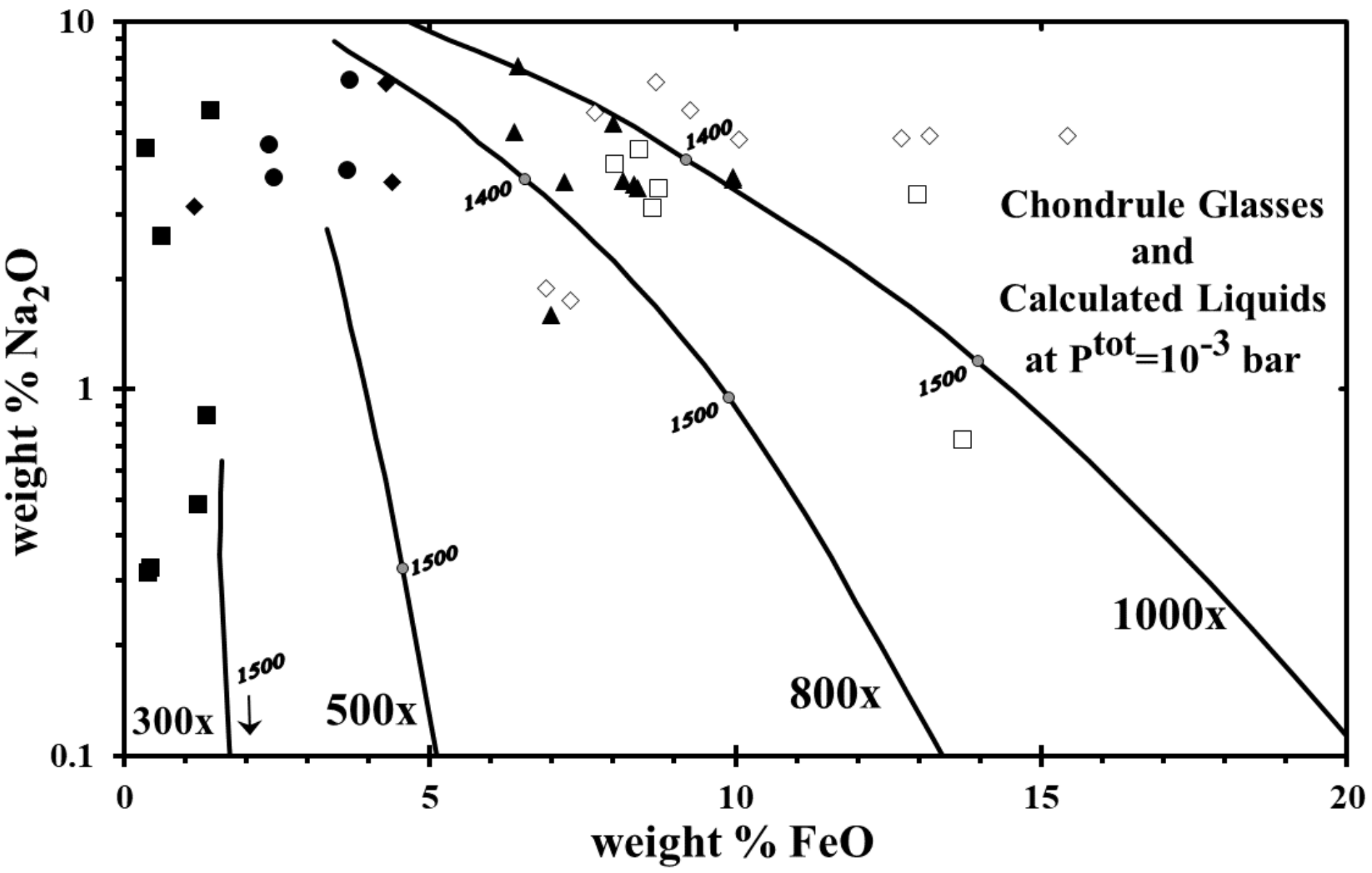

**Figure 19:** Glass compositions in chondrules, with paths of condensate liquids for 300, 500, 800 and 1000x C1 dust enrichment, which terminate at their solidi. Data as in Fig. 18.

The liquid paths in Fig. 18 suggest that chondrule glasses with high FeO + alkali oxide contents could have equilibrated with a highly dust-enriched gas at $10^{-3}$ bar. The glass compositions observed in Type IA chondrules, however, contain much less MgO + $SiO_2$ than any FeO- or alkali-bearing liquids in equilibrium with dust-enriched vapor. In these particular chondrules, it is mostly $Na_2O$ content which pulls the liquid composition off the CMAS join, into the interior of the triangle. This can be seen in Fig. 19, which shows glass compositions of the same chondrules, with paths of liquid composition for four dust enrichment factors superposed. Type IA chondrules fall closest to the y-axis. In the case of 1000x enrichment at $10^{-3}$ bar, the FeO contents of calculated liquids increase to >25 wt% while $Na_2O$ <0.1 wt%, then decrease with decreasing temperature. Past the peak in FeO content, Na begins to condense into liquids, and $Na_2O$ concentrations increase steeply near the solidus, at which points the curves in Fig. 19 terminate. Prior to Na condensation, $P_{Na}$ is ~2.7x$10^{-6}$ bar at a dust enrichment of 1000x, almost three times higher than that at 300x. Because of this and the slightly higher $f_{O2}$ in the most dust-enriched systems, 800x and 1000x, a significant fraction of the Na condenses into the liquid above 1400K. As a result, below 1400K, $Na_2O$ contents of the calculated liquids at dust enrichments of 800x and 1000x are in the range of those of many of the chondrules shown, and the calculated $P_{Na}$ is very similar for all dust enrichments, varying from ~3.2x$10^{-7}$ bar at a dust enrichment of 800x to ~4.8x$10^{-7}$ bar at 1000x at 1350K. The most important reason why the $Na_2O$ contents of the liquids at the highest dust enrichments reach the levels found in chondrules while those at the lowest dust enrichments, 300x and 500x, do not is the persistence of the former to lower solidus temperatures, ~1300K *vs.* ~1400K (Fig. 8). As discussed above, this is due to higher FeO and $Na_2O$ contents of the liquids in the more dust-enriched systems at 1400K. At constant dust enrichment, our calculations show that maximum alkali contents of the liquids decrease substantially with decreasing $P^{tot}$. Even at $P^{tot}$ as high as $10^{-4}$ bar and a dust enrichment of 1000x, for example, the maximum $Na_2O$ content of the liquid is at least a factor of 5 smaller than at $10^{-3}$ bar and the same dust enrichment. Except for the Type I chondrule glasses with $Na_2O$/FeO wt. ratios > 2.0, and Type II chondrule glasses with very high FeO *and* $Na_2O$ contents, all the chondrule glass compositions examined here could represent silicate liquids in equilibrium with dust-enriched vapor within 200 degrees of their solidus temperatures at $P^{tot}$ > $10^{-4}$ bar. Even if the precursors of the most FeO- and $Na_2O$-rich chondrules were melted in a gas enriched 1000x by dust, their liquids would have lost sodium by evaporation or FeO by reduction, assuming that they were hot for a long enough time to have equilibrated with the gas.

## 5. CONCLUSIONS

Condensation of systems sufficiently enriched in dust of C1 composition to yield ferromagnesian silicates with molar FeO/FeO + MgO ratios of 0.1 to 0.4 at temperatures above 1200K also produces copious molten silicate. The distribution of Fe between metal, silicate and sulfide in specific classes of ordinary chondrites can be produced by high-temperature condensation at specific dust enrichments. While a rigorous thermodynamic model at last shows how direct condensation of silicate liquid can occur within the range of $P^{tot}$ thought to have existed in the inner part of the solar nebula, the compositions of the partially molten condensates so formed do not match the compositions of Types I and II chondrules in important ways. Such chondrules are thus secondary objects that did not form by direct condensation. Nevertheless, if chondritic matter owes its oxidation state to condensation of dust-enriched systems, the present work shows the sequence of condensation, details of the condensation reactions and the evolution of the compositions of solid and liquid solution phases that may be relevant to the formation of chondrites and the precursors to chondrules. The present work also gives physicochemical conditions capable of stabilizing against evaporation silicate liquids of specific compositions, some of which are similar to the compositions of some chondrule glasses, in cosmic gases at low nebular pressures.

*Acknowledgements* - The authors extend thanks to J. Valdes, S. Champion, D. Archer and G. Miller for technical assistance, and to M. S. Ghiorso and R.O. Sack for providing advice and the MELTS code. Critical reviews by J.R. Beckett and an anonymous reviewer were very helpful. Material support for this project was through NASA grants NAGW-3340 and NAG5-4476.